%% file: paper.tex
\documentclass{tex/src/aa}

\input{tex/build/macros/project.tex}

\input{tex/src/preamble-maneage.tex}

\input{tex/src/preamble-project.tex}

\title{\projecttitle}
\subtitle{}

\author{
  Sepideh Eskandarlou\inst{\ref{CEFCA}}
  \and
  Mohammad Akhlaghi\inst{\ref{CEFCA},\ref{UA}}
  \and
  Johan H. Knapen\inst{\ref{IAC},\ref{ULL}}
  \and
  Carlos L\'opez-Sanjuan\inst{\ref{CEFCA},\ref{UA}}
  \and
  Ra\'ul Infante-Sainz\inst{\ref{CEFCA}}
  \and
  Helena Dom\'inguez S\'anchez\inst{\ref{CEFCA}, \ref{IFCA}}
  \and
  Zahra Sharbaf\inst{\ref{IAC},\ref{ULL}}
  \and
  H\'ector V\'azquez Rami\'o\inst{\ref{CEFCA},\ref{UA}}
  \and
  Juan Antonio Fern\'andez Ontiveros\inst{\ref{CEFCA},\ref{UA}}
  \and
  C\'esar I\~niguez García\inst{\ref{CEFCA}}
  \and
  Tamara Civera Lorenzo\inst{\ref{CEFCA}}
  \and
  David Jos\'e Muniesa Gallardo \inst{\ref{CEFCA}}
  \and
  Paula R.T. Coelho\inst{\ref{USP}}
  \and
  Alessandro Ederoclite\inst{\ref{CEFCA},\ref{UA}}
  \and
  Jesus Varela\inst{\ref{CEFCA}}
  \and
  Fran Jim\'enez-Esteban\inst{\ref{CSIC}}
  \and
  A. Javier Cenarro \inst{\ref{CEFCA},\ref{UA}}
  \and
  Antonio Mar\'{\i}n-Franch\inst{\ref{CEFCA},\ref{UA}}
  \and
  Renato A. Dupke\inst{\ref{ON},\ref{MU}}
  \and
  Mariano Moles\inst{\ref{CEFCA}}
  \and
  Carlos Hern\'andez-Monteagudo\inst{\ref{IAC},\ref{ULL}}
  \and
  Rahna P.T.\inst{\ref{CEFCA}}
  \and
  David Crist\'obal-Hornillos\inst{\ref{CEFCA}}
  \and
  Jailson Alcaniz\inst{\ref{ON}}
  \and
  Laerte Sodr\'e Jr.\inst{\ref{USP}}
  \and
  Raul E. Angulo\inst{\ref{DIPC},\ref{ikerbasque}}
}

\institute{Centro de Estudios de F\'{\i}sica del Cosmos de Arag\'on (CEFCA), Plaza San Juan 1, 44001 Teruel, Spain\label{CEFCA}
  \and Unidad Asociada CEFCA-IAA, CEFCA, Unidad Asociada al CSIC por el IAA y el IFCA,
  Plaza San Juan 1, 44001 Teruel, Spain\label{UA}
  \and Instituto de Astrof\'{\i}sica de Canarias, La Laguna, 38205, Tenerife, Spain\label{IAC}
  \and Departamento de Astrof\'{\i}sica, Universidad de La Laguna, 38206, Tenerife,
  Spain\label{ULL}
  \and Instituto de F\'{\i}sica de Cantabria (IFCA), CSIC-Univ. de Cantabria, Avda. los Castros, s/n, E-39005 Santander, Spain\label{IFCA}
  \and Observat\'orio Nacional - MCTI (ON), Rua Gal. Jos\'e Cristino 77, S\~ao Crist\'ov\~ao,
  20921-400 Rio de Janeiro, Brazil\label{ON}
  \and University of Michigan, Department of Astronomy, 1085 South University Ave., Ann
  Arbor, MI 48109, USA\label{MU}
  \and Instituto de Astronomia, Geof\'{\i}sica e Ci\^encias Atmosf\'ericas, Universidade de
  S\~ao Paulo, 05508-090 S\~ao Paulo, Brazil\label{USP}
  \and Centro de Astrobiolog\'{\i}a, CSIC-INTA, Camino bajo del castillo s/n, E-28692, Villanueva de la Can\~ada, Madrid, Spain\label{CSIC}
  \and Donostia International Physics Centre (DIPC), Paseo Manuel de Lardizabal 4, 20018
  Donostia-San Sebastián, Spain\label{DIPC}
  \and IKERBASQUE, Basque Foundation for Science, 48013, Bilbao, Spain\label{ikerbasque}
}

\date{Received July 22, 2025; accepted October 12, 2025}

\abstract
 {
   Photometric surveys require precise point spread function (PSF) characterization, as it varies across filters and is crucial for accurate photometry and low surface brightness (LSB) studies, such as galaxy halos, tidal features, extended emission from the circum-galactic medium, and intra-cluster light.
   However, the small PSF size provided by default pipelines suits only barely resolved objects, making it difficult to analyze regions near bright stars (rendering those regions unusable).
 }
 {
   We aim to demonstrate the feasibility of subtracting the extended PSF from each J-PLUS DR3 exposure prior to sky subtraction, an approach that has not yet been explored in wide surveys, and to enable a comprehensive analysis of its behavior across detector position and time.
}
{
  To build an extended, non-parametric PSF, three different ranges of stars are selected to create the central, middle, and outer regions in exposures within $\sim2.5$ hours of the target.
  These components are then combined to generate a final PSF for each exposure and filter, spanning $\psfsurfacebrightness$\,mag\,arcsec$^{-2}$ in surface brightness and $\psfsize$\,arcmin in radius in the broad bands.
}
{
  In narrow-band filters, the J-PLUS PSF exhibits two rings, whereas in broad-band filters, only one ring is observed.
  Additionally, the position of the ring shifts with filter wavelength: as the filters become redder, the ring radius increases.
  We find that the precision of sky subtraction can be greatly improved with a PSF-subracted image and that out to a radius of $\psfsize$\,arcmin, there is no significant variation in the extended PSF observed as a function of time or position in the field of view.
  The radial profile of NGC 4212 (which is close to a star) is also studied before/after PSF-subtraction as a demonstration of the effect.
  We developed a novel method to determine the central coordinates of saturated stars, and classify stars without using Gaia magnitudes.
  Additionally, mirror reflections are automatically detected and masked.
  Furthermore, in combining different stars and various components of the PSF, we avoided the use of a fixed radius by introducing a new method that does not depend on radial measurements.
}
{
  Accurate characterization of the extended PSF and its subtraction improves sky subtraction, increases the effective area of the survey by about 10\%, and enables the study of extended large LSB features in wide area surveys like J-PLUS.
  Our pipeline is published as free software (GNU GPLv3) an can be customized to other surveys such as J-PAS, where its impact will be even greater due to its depth.
  This paper is fully reproducible and produced from Commit \texttt{\projectversion}.
}


\begin{document}
\maketitle

\section{Introduction}
The blue sky indicates how far light can scatter from the position of an astronomical light source (the Sun!).
In optical imaging systems, the light from a point source is concentrated into a ``point'' that is strongly ``spread'' (or scattered) across the detector by the atmosphere, telescope and camera, as a 2D ``function''.
This is referred to as the point spread function (PSF).

However, the standard pipelines of wide-field surveys only provide the PSF to very small sizes (for example 10 arcsec).
While this is useful for the study of barely resolved source (for example high redshift galaxies), it does not allow the removal of the scattered light which has been produced in the image by the bright sources.
Estimating the extended PSF (up to several arcminutes) is crucial for an un-biased study of many aspects of galaxy evolution.
For instance, the drop in completeness of barely resolved objects near bright stars, colors of galaxies, radial profiles and truncation's, measurement of the sky, halo shape, tidal features, dwarf galaxies, circum-galactic medium (CGM), and intra-cluster light (ICL).
See \citet{knapen17} for a review.

Bright stars are a limiting factor for large imaging survey because their light scatters to relatively large distances and faint surface brightness levels; becoming more significant as the survey gets deeper.
The influence of scattered light is so prominent that many projects are forced to mask large portions of their field that are affected.
For instance in the Hyper-Suprime-Cam Strategic Subaru Proposal (HSC-SSP) up to $20\%$ of the total area has been masked \citep[see fig.~15 of][]{coupon18} due to brighter stars.
In the Javalambre Photometric Local Universe Survey (J-PLUS) DR3\footnote{\url{https://www.j-plus.es/datareleases/data_release_dr3}}\citep{cenarro19}, approximately $10\%$ of the survey area is lost due to bright stars (less than in HSC-SSP because the J-PLUS is shallower).

The accurate characterization of, and accounting for, the extended PSF is important for many aspects of imaging-based astrophysical research and has been shown to lead to better and more reliable results in areas of study ranging from point sources \citep{anderson00, mighell05, jimenez15, nardiello22} to gravitational lensing \citep{gillis20, nardiello22}.
But it is critical in studies of extended emission around astrophysical objects \citep[e.g.][]{sandin15}.
For example, scattered light can create artificial structure in galaxies that can closely mimic real structure, when it is azimuthally smooth, as a galaxy outer disk or halo \citep{michard02, idiart02, trujillo16, peters17}.
The imaging of tidal structures around galaxies can also be severely affected by scattered light \citep{vandokkum19, martinz08, Lanzetta23}, as can the study of thick disks in galaxies \citep{martinexlomnilla19, comeron18}.
Also, the outer parts of planetary nebulae (PNe) are affected by scattered light \citep{middlemass89}.

However, many surveys just characterize the central few arcsec region of the PSF, which does not allow the appropriate removal of the scattered light produced by bright sources in or even outside the imaged area \citep{nacho01a, nacho01b}.
Quantifying the extended PSF (up to several arcmin) is therefore crucial for the correct detection of the faint features which highlight important aspects of galaxy evolution and many other galactic sources.

As one pushes the detection limit to fainter surface brightness levels, or observes closer to the disk of the Milky Way, the extended PSF becomes one of the most significant limiting factors.
Such that their increased presence is one of the factors in a survey's design.
The problem of the PSF is not limited to ground-based surveys.
For example fig.~8 of \citet{borlaff22} shows the expected surface brightness of the PSF on average in the Euclid survey, and fig.~3 of that paper illustrates how much of the Euclid survey area will be affected by stars.

The common solution of masking the bright stars is not the optimal way to deal with this issue, because it leads to the loss of significant sky coverage and contiguity in a survey.
Furthermore, the extended PSF is usually much larger than the masked area (which is primarily designed for point-source photometry).
Historically, \citet{moffat69} and \citet{nacho01b} noticed that a simple Gaussian is not a good parametric fit and defined a new function for this purpose.
\cite{king71} used the Sun to construct an extended PSF that reaches out to six degrees.
\citet{devacouleurs58} used Jupiter to measure the PSF out of five degrees.
While the Sun or Jupiter can be used to construct a degree-level PSF, we can not use such a PSF.

A more robust solution for this problem is to accurately identify the extended PSF and subtract it from bright stars \citep[see e.g.,][] {mihos13, sandin14, sandin15, roman20, infantesainz20, liu22, nunez23, nafise25}.
There are generally two approaches to constructing the PSF: \citet[hereafter I20]{infantesainz20} uses the coadd of thousands of star images (where other sources are masked) to construct a non-parametric PSF out to $8$ arcmin in radius.
On the other hand \citet{liu22} uses a parametric Moffat and multi-power-law fit to each star, which is only good when the PSF is perfectly circular symmetric, without rings, spikes or ghosts (artifacts that are common in large reflective telescopes).

The structure of the PSF is not fixed and changes with various parameters.
For example, the outer part of the PSF, also known as the aureole, is affected by diffusion and reflection within the instrument \citep{hasan95}.
\cite{michard02} selected stars and measured the variation of the outer wings of the PSF with color, concluding that the atmospheric seeing did not change, while the PSF wings depend on the filter and on the time elapsed since the coating of the mirror.
\citet{xin18} modeled the SDSS PSF, studied the seeing profile, and found that the PSF changes with wavelength and time.
Their PSF was constant at a time scale ranging from 5 to 30 minutes.
\cite{liu22} showed how accumulated dust on lenses affected the extended PSF.

In this paper we aim to construct, subtract and study the extended PSF of the J-PLUS DR3 survey.
Given that J-PLUS has spikes and rings in its PSF, parametric methods are not useful.
Therefore, we use the concepts introduced by I20 and greatly improve upon them for the much more complex data from J-PLUS.
Furthermore, whereas I20 produced PSFs using the final coadded size of each T80Cam image is $2$ deg$^2$ with a pixel scale of $0.55$ arcsec pix$^{-1}$; therefore the images are $9200 \times 9200$ pixels \citep{Antonio15}.

The design of the J-PLUS filters is optimized for sampling the spectral energy distribution (SED) of Milky Way (MW) stars and nearby galaxies \citep{cenarro19}.
The camera is described in detail by \cite{Antonio15}, the survey design and the scientific aims by \cite{cenarro19}.
The photometric calibration and the third data release is described by \cite{carlinhos23}.

In this work, we use $\numtilespaper$ single-exposure images from the third J-PLUS data release.
This is 10\% of the total amount of imaging released so far.
The focus of this paper is to introduce the pipeline, analyze the J-PLUS PSF and review some of the advantages in a subset of the tiles, not the whole survey.
We hope to extend the application of this pipeline to future J-PLUS data releases, integrating it across all exposures during the data reduction process.
Data reduction generally involves several steps, including dark subtraction, flat-fielding, astrometric calibration, and sky subtraction.
Our aim is to incorporate PSF subtraction into the data reduction sequence, placing it before the sky subtraction stage.

The construction of the PSF requires a prior knowledge of the location and some properties of the stars.
For those, we use Gaia Data release 3 \citep{gaiadr3}.

\section{Building the extended J-PLUS PSF}
\label{sec:creatpsf}

\begin{figure}[]
  \begin{center}
  \ifdefined\makepdf%
    \tikzsetnextfilename{fig-flow-chart}%
    \input{tex/src/fig-flow-chart.tex}%
  \else
    \includegraphics[width=\linewidth]{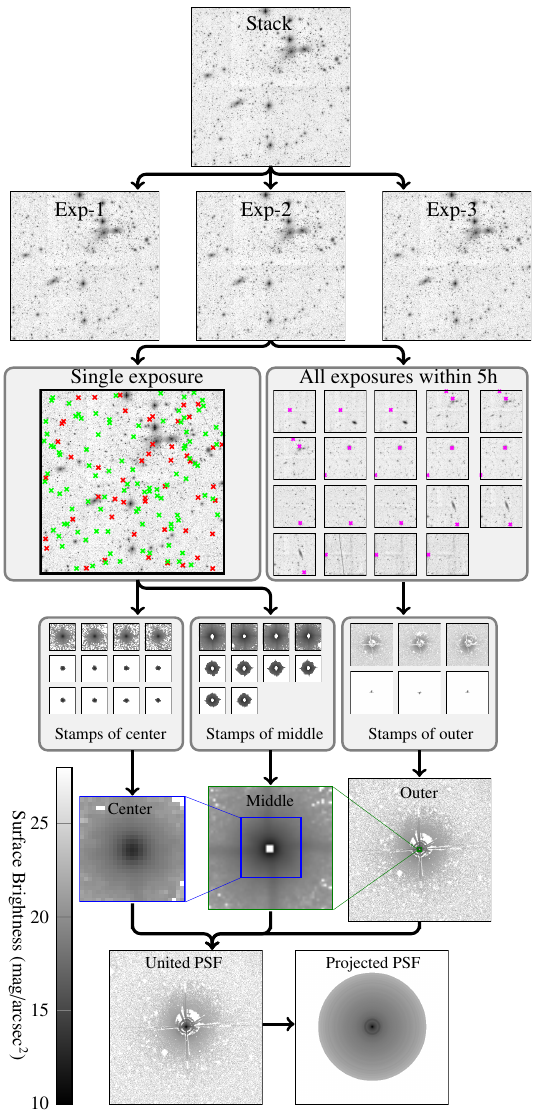}
  \fi

  \end{center}
  \caption{\label{fig:flowchart}Flow chart of extended the PSF construction for a J-PLUS tile.
    All images are scaled to the color-bar at the bottom.
    The top row shows the coadded image of three exposures which are shown in second row and the second row shows the three exposures that were used to build it.
    The third row shows the exposure(s) used for the various parts of the PSF in the second exposure of the second row: the single exposure itself for the center and middle parts and all exposures $\pm2.5$ hours before/after the exposure for the outer part.
    The marked green and red stars are used for creating central (Sect.~\ref{sec:createcenter}) and middle parts (Sect.~\ref{sec:createmiddle}), respectively.
    The stars under pink crosses (on the right) are used to create the outer PSF (Sect.~\ref{sec:constructout}).
    The fourth row presents a selection of the star stamps, with other sources masked, while the fifth row displays the coadded stamps.
    These are subsequently combined to construct the ``Unified PSF'', shown in the final row.
    Its 1D radial profile up to {$\psfsizeouter$} arcmin (in radius) is then projected into the circular ``Projected PSF''.
    The surface brightness of the united PSF (from the central pixel to outer) changes about {$\psfsurfacebrightness$}\,mag\,arcsec$^{-2}$ (See Sect.~\ref{sec:finalpsf}).
    }
\end{figure}

As summarized in the introduction, we adopt a non-parametric approach to constructing the extended PSF, based on the previous work by I20.
But while I20 constructed a single PSF for the whole SDSS, this work constructs it for every exposure; many other improvements have been introduced that are further elaborated in the text.
Fig~\ref{fig:flowchart} summarizes all the steps for J-PLUS DR3 tile $\flowcharttilerefid$ in the {\flowchartfilter} filter, which contains the Coma cluster (just under the label on the top image).

The overall approach (as a summary of the subsections below) is that non-saturated stars in the image are used to construct the very central few pixels of the PSF, while the outskirts of the brightest stars are used for the outer part of the PSF, out to distances of several arcmin.
Intermediate-brightness stars are used to connect the middle and center part of the PSF, thus creating a profile which spans up to several arcmin in radius and covering (from brightest to faintest part) up to {$\psfsurfacebrightness$} mag/arcsec$^2$ in surface brightness.
These extended PSFs are created without any assumptions on the functional form of the PSF, and are thus robust to any peculiarities that are present in many optical systems.
The subsections below we describe the intricacies of constructing each part of the PSF.

For PSF construction, an internal sky estimate is derived and subtracted using the \textsf{NoiseChisel} program \citep{gnuastro}.
This intermediate step should not be confused with the formal sky subtraction performed later in the data reduction pipeline, as the \textsf{NoiseChisel} parameters may differ between the two stages.

As shown in Fig~\ref{fig:flowchart}, the different components of the PSF are constructed by generating stamps around known stars.
These stamp images were produced with the \texttt{ast\-script\--psf\--stamp} tool, as described in Sect. 10.8.3 of \citet{gnuastrobook}.
Each stamp is a cropped image centered—sub-pixel aligned when necessary—on the given coordinates, with all foreground and background sources masked.
The stamp size is consistent across filters and is manually chosen based on broad-band filters, as detailed in the following subsections.

\subsection{Selection of stars to use}
\label{sec:starselection}
Individual stars are necessary at very different magnitudes to construct the various parts of the extended PSF.
In particular, we choose stars with a good parallax from Gaia DR3: their parallax should be greater than three times the parallax error.
This is done to avoid confusion with quasars for the central and middle parts.
The selected stars must be sufficiently isolated from nearby bright sources to minimize contamination.
Isolation is defined using the stamp size as the reference radius: a star is considered isolated if no brighter source is detected within this radius.
Such stars are then consistently used to construct the different components of the PSF.
Before using the images of each star, some pre-processing was necessary:
Based on the saturation and non-linearity limit in J-PLUS, saturated pixels were masked in all the exposures used.
Based on the saturation and non-linearity limits in J-PLUS, saturated pixels were masked in all exposures.

I20 explored the possibility of using Gaia broad-band magnitudes for their selection and sorting of the brightest stars.
They found that Gaia is incomplete for objects brighter than 7th magnitude, and no stars with magnitude below 1.7 mag are present.
Furthermore, stars may be brighter or fainter in specific narrow bands, and the Gaia magnitudes of such stars may not match their actual brightness within a narrow/medium-band image.

The current work is a proof of concept for the Javalambre Physics of the Accelerating Universe Astrophysical Survey \citep[J-PAS;][]{benitez14,minijpas} survey.
J-PAS uses $56$ narrow-band filters and will thus suffer this narrow band sorting issue much strongly.
Due to the varied properties of stars across 56 narrow-band filters, it is inappropriate to use Gaia broad-band magnitudes for sorting stars.
Given all the issues above, we chose not to use Gaia magnitudes to select and sort the stars.

\subsection{Constructing the central part of the PSF}
\label{sec:createcenter}
Non-masked (due to saturation and non-linearity) stars in each exposure are selected to construct the central part of its PSF.
Gnuastro's \textsf{Segment} \citep{akhlaghi19} and \textsf{MakeCatalog} \citep{akhlaghi19b} programs are used to mask other sources and find these stars.
Additionally, all those with any bright stars or galaxy in the neighborhood were rejected.
The stars were then sorted based on the mean value of the brightest three pixels and the brightest {$\psfnumbercenter$} stars were selected.
As mentioned in Sect.~\ref{sec:starselection} this allows us to not use Gaia's magnitudes.
The number of stars is high enough to reach an adequate signal-to-noise ratio (larger than 3) in the central {$\psfsizecenterpaper$} arcmin stamp of the PSF, the stamp size is constant in the pipeline but can be adjusted by the user.
The positions of the stars thus selected are marked with a green cross in the ``single exposure'' box of Fig.~\ref{fig:flowchart}.

The ``stamps of center'' box in Fig.~\ref{fig:flowchart} displays the image stamps of several non-saturated stars used to construct the central part of the composite PSF.
All pixels with a signal-to-noise ratio less than three are set to NaN (Not a number).
Only the four brightest stars out of the {$\psfnumbercenter$} selected stars are not masked and fully considered in the stacking procedure to construct the PSF.
The number four is chosen arbitrarily only to ensuring sufficient signal in the connecting region with the next part of the PSF, which will cover the majority of these pixels.
This procedure ensures that fainter stars are excluded from the outer regions, thereby preventing the addition of unwanted noise.

\begin{figure}[!t]
  \begin{center}
  \ifdefined\makepdf%
    \tikzsetnextfilename{fig-radial-profile}%
    \input{tex/src/fig-radial-profile.tex}%
  \else
    \includegraphics[width=\linewidth]{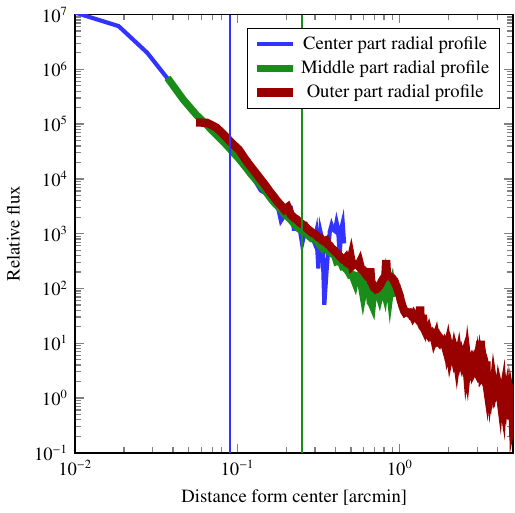}
  \fi

  \end{center}
  \captionof{figure}{\label{fig:radone}
    Radial profiles of the coadds used for different PSF components (fifth row of Fig.~\ref{fig:flowchart}).
    The blue, green, and red profiles correspond to the central, middle, and outer regions, covering radii of {$\psfsizecenterpaper$}, {$\psfsizemiddlepaper$} and {$\psfsizeouter$} arcmin (in radius), respectively.
    Vertical lines indicate the points where these regions are joined.}
\end{figure}

The next step in creating the central part of the PSF is to coadd the stamps of all individual stars.
But this can only be done after they are all normalized to the same flux.
Most previous approaches such as I20 used a specific radius to obtain the normalization value in the specific radius range.
Similarly, \citet{amir24} demonstrated that selecting an appropriate normalization radius is particularly challenging in large pipelines and plays a crucial role in creating the extended PSF for the Vera C. Rubin Observatory’s Legacy Survey of Space and Time (LSST).
J-PLUS has even more complex issues because of its number and diversity of filters and the goal of this work (to have an extended PSF for each exposure).

For the normalization, we consider the brightest star as a reference.
We use Gnuastro’s \texttt{ast\-script\--psf\--stamp}, which applies a consistent masking strategy across all steps described in this paper.
This tool crops both the reference and target stars, masks extraneous objects, and centers the images with sub-pixel precision.
Then, all pixels fainter than a signal-to-noise ratio of {$\normsigmaout$} are masked and we find the normalization value by dividing the sum of the remaining pixels of each star by that of its reference star.
As all stars are centered in the central pixel of each stamp, this method automatically leads to the correct normalization value for each star but without having to manually specify a certain radius.

After normalization we coadd all the stamps by taking the mean after $3\sigma$-clipping to construct the central part of the PSF (shown in the left panel of the fifth row of Fig~.\ref{fig:flowchart}).
The radial profile \citep[using Gnuastro's \texttt{ast\-script\--radial\--profile; see}][]{raul24} of the central part of the PSF is shown in Fig.~\ref{fig:radone} as the blue curve (smallest radii).

\begin{figure}[!t]
    \begin{center}
  \ifdefined\makepdf%
    \tikzsetnextfilename{fig-sat-obj-coord}%
    \input{tex/src/fig-sat-obj-coord.tex}%
  \else
    \includegraphics[width=\linewidth]{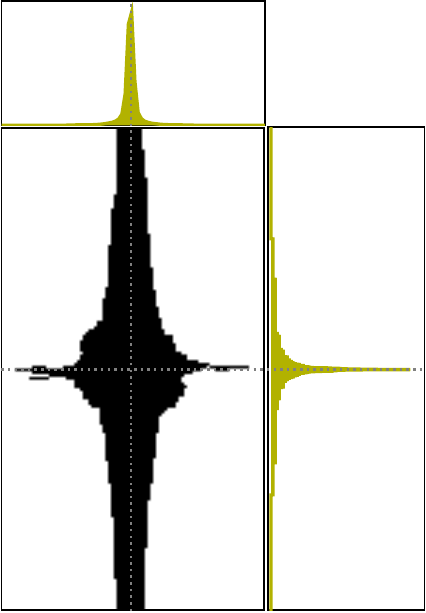}
  \fi

    \end{center}
    \captionof{figure}{\label{fig:objcoord}
      Finding the center of saturated stars from the bleeding pixels (black in bottom-left panel).
      After collapsing the pixels along X and Y axis, the peak of the distribution shows the center coordinate of the saturated star.}
\end{figure}

\subsection{Constructing the middle part of the PSF}
\label{sec:createmiddle}
The method for constructing the middle part of the PSF is quite similar to that used for the central part.
The middle part of the PSF for each tile is also composed of the stars from that tile, but here we use the faintest stars affected by non-linearity or saturation\footnote{Saturated pixels occur when a sensor reaches its maximum charge capacity, causing photons to leak into neighboring pixels.
  Non-linear pixels occur at fainter pixel values where the electron count does not increase at the same rate as the input flux.
  Therefore, these are not considered part of the PSF.}.
Information on how these pixels are identified and masked can be found in Sect.~2.3.2 (``Saturated pixels and Segment’s clumps'') of \citet{gnuastrobook}.

We then match the resulting catalog with Gaia coordinates and select those stars whose parallaxes are greater than three times of the parallax error.
The stars are then sorted by their signal-to-noise ratio and the {$\psfnumbermiddle$} faintest saturated stars are selected for stacking.
These stars are marked with red crosses in the ``Single exposure'' box of Fig.~\ref{fig:flowchart}.
The choice of {$\psfnumbermiddle$} is arbitrary and defined by the user, and the actual number used may vary across different tiles.

The remaining procedure follows the same steps as in the construction of the central part of the PSF.
Adequate faint, saturated stars are selected, the sky is subtracted, and a stamp image is generated for each (Fig.~\ref{fig:flowchart}, middle panel of the fourth row).
The stamp size is fixed at {$\psfsizemiddlepaper$} arcmin across the pipeline, although it can be modified by the user.
During the stamp creation, all extraneous objects are masked.
Pixels with a signal-to-noise ratio lower than three are then removed, with the exception of the four brightest stars.
As mentioned in the previous section, retaining only high signal-to-noise pixels ensures that faint sources do not contribute additional noise, particularly in the outskirts.
The resulting stamps are normalized following the method outlined earlier and then co-added to form the composite middle part.
The green curve in Fig.~\ref{fig:radone} displays the radial profile of the middle part of the PSF.
It covers {$\psfsizemiddlepaper$} arcmin of the total PSF.

\begin{figure}[!t]
  \begin{center}
  \ifdefined\makepdf%
    \tikzsetnextfilename{fig-sort-bright-sat-star}%
    \input{tex/src/fig-sort-bright-sat-star.tex}%
  \else
    \includegraphics[width=\linewidth]{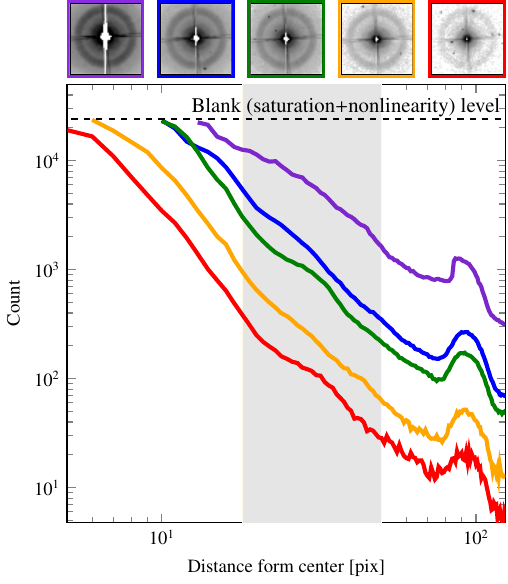}
  \fi

  \end{center}
  \captionof{figure}{\label{fig:sortstars}
    Sorting the brightest stars in an image by magnitude using flux at a certain radius.
    Colored radial profiles correspond to stars in the top of the image (same same color border).
    Stars are sorted based on flux within the vertical gray region (which is just beyond the saturated radius of the brightest star).
  }
\end{figure}

\begin{figure*}[!t]
  \begin{center}
  \ifdefined\makepdf%
    \tikzsetnextfilename{fig-sat-ghost}%
    \input{tex/src/fig-sat-ghost.tex}%
  \else
    \includegraphics[width=\linewidth]{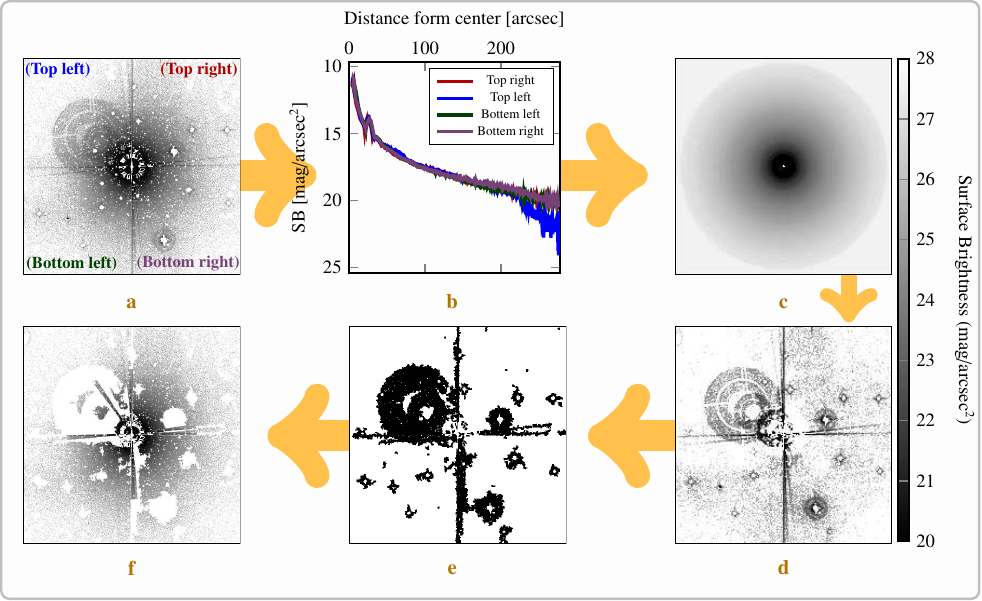}
  \fi

  \end{center}
  \caption{\label{fig:satghost}
    Flowchart of identifying and masking internal reflections (ghosts).
    The star's image (panel a) is divided into quadrants by azimuthal angle (0-90,90-180,180-270,270-360 degrees).
    The radial profile of different angles are found (in blue, red, green, and purple) and shown in panel b.
    The quadrant which has less signal-to-noise ratio (in this case the violet one) is chosen to create a 2D circular projection of that profile (panel c).
    The 2D projection is subtracted from the input, highlighting all problematic signal (including ghosts and wings of fainter stars) in panel d.
    We detect all of those low signal-to-noise with NoiseChisel in panel e and then mask them in panel f.
    The pixels that belong to spikes are re-inserted, as described in the text.
  }
\end{figure*}

\subsection{Constructing the outer part of the PSF}
\label{sec:constructout}
The construction of the outer part of the PSF is the most novel aspect of this work, and also the most complicated.
This is because the outer part of the PSF can only be observed around the brightest stars.
The ``All exposures in $\psfdeltahour$ hours'' box in Fig.~\ref{fig:flowchart} outlines the construction steps of the outer part of the PSF, using the brightest stars observed.
As the number of bright stars in each tile is limited and we need to increase the signal-to-noise ratio in the outer part of the PSF, we use exposures taken with the same filter within a $\psfdeltahour$-hour interval of each exposure.
If there is a night with fewer valid observation, we use only the images taken during that time.
Within these five hours, the telescope pointing can change, we have studied the behavior of the outer part of the PSF in different times in Section \ref{sec:dependontime}.
In all stages of constructing the outer PSF, as in the central and middle regions, the sky is subtracted and extraneous objects are masked, following the methodology described earlier.

The outer part of the PSF, spanning a radial range of {$\psfsizeouter$} arcminutes, is constructed using the brightest saturated stars.
A crucial, but by no means trivial, step is to determine their central position in the image.
We could not use Gaia for the central positions of these stars because it is not complete for the brightest stars.

To find their centers we use the bleeding area of these stars (which is nearly symmetric for the T80Cam detector; the slight asymmetry has no significant effect) as shown in Fig.~\ref{fig:objcoord}.
We crop each bleeding region, setting all saturated/bleeding pixel values to $1$ and all other pixel values to NaN.
The image is then collapsed along the horizontal/vertical axes producing a 1D column that shows how many masked pixels were in each.
The distribution along the axes can be seen in the top and right side of the bleeding region of Fig.~\ref{fig:objcoord}.
The maximum values in X and Y axes are shown with a gray dashed line, defining the coordinates of the center of the saturated star.
Note that this method is specifically designed for use with the J-PLUS detector and is only generalizable to detectors which have a symmetric bleeding pattern.

After defining the central coordinates of the saturated stars, the next critical problem with very bright stars is the selection and sorting of these bright stars.
If this is not carefully accounted for and we stack stars with a wide range of brightness, an artificial truncation will occur in the outer part of the PSF caused by the noise in the outer part of images of the fainter stars.
Like before, due to the diversity in the J-PLUS filters, we cannot simply use the Gaia broad-band magnitudes and due to the heavy saturation/bleeding in their centers, a simple read-out of the clumps from \textsf{Segment} (as in the middle part) is not enough.
Aperture photometry is also not reliable because of two factors: the bleeding area can be different and we need stars that are also located on an edge of an image.

To sort the bright stars by flux, the stars with a saturated area larger than $\satareainpix$ pixels are first selected.
For clarification, saturated and non-linear pixels are recorded as NaN, so their area is determined from the contiguous NaN regions, and their original pixel values are excluded from the analysis.
In contrast, pixels identified as spikes are included in the analysis, because saturated pixels are not part of the PSF, whereas spikes are.
This procedure is applied to all exposures taken within the $\psfdeltahour$-hour interval of each target.
We crop the images of these stars to a size of $\sortsatstarwidth$ pixels, as this dimension nearly includes the stellar rings (which will be discussed later but are not considered in this analysis) and remains sufficiently distant from the mirror reflection region.
Finally, their radial profiles are generated.
The cropped images and the radial profiles of some of these stars are shown in Fig.~\ref{fig:sortstars} (note that in all images, darker pixels indicate higher brightness), where we show the saturation and non-linearity level of J-PLUS's reduced exposure with a horizontal dashed black line.
The brightest star is the one for which the smallest non-blank (saturation+nonlinearity level) radius is the most distant from the center, the purple one in the figure (note that darker pixels are brighter in this paper's images).
After identifying this radius among all the stars, and in order to avoid saturated pixels that might appear around the bleeding area, we selected a point 5 pixels beyond the smallest non-blank radius as the minimum boundary of the light gray area.
This value was then multiplied by $\outmultipleradius$ to determine the maximum extent of the light gray area in Fig.~\ref{fig:sortstars}.
The flux within the interval is summed and used for sorting stars by brightness.
The interval does not terminate when the purple profile concludes, as we excluded the star's ring to avoid confusion due to the varying brightness of the ring among different stars.
Among all the exposures shown in Fig.~\ref{fig:flowchart}, we sorted and selected stars based on this methodology.
All the stars used to construct the outer part of the PSF are marked with pink crosses in Fig.~\ref{fig:flowchart}.

Another important obstacle in the construction of the outer part of the extended PSF are the internal reflections (also known as ghosts).
Such reflections occur at faint levels in the vicinity of bright stars due to light bouncing around in the instrument/telescope system and are very hard to model \citep{slater09}.
They do not originate from diffraction (and are therefore not part of the PSF).
Therefore, we need to mask them for each star when constructing the outer part of the PSF.

\begin{figure*}[!t]
  \begin{center}
  \ifdefined\makepdf%
    \tikzsetnextfilename{fig-psf-subtract}%
    \input{tex/src/fig-psf-subtract.tex}%
  \else
    \includegraphics[width=\linewidth]{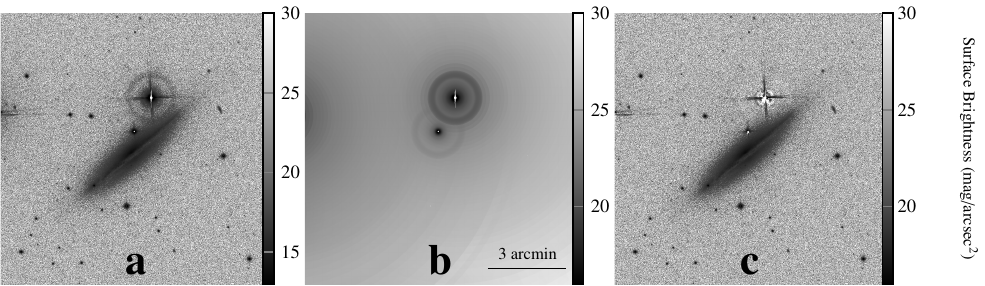}
  \fi

  \end{center}
  \caption{\label{fig:psfsub}
    NGC 4212 in the J-PLUS rSDSS filter before (panel a) and after (panel c) PSF subtraction.
    Panel b displays the PSF that was subtracted from panel a to create panel c, which also displays the residuals.
      In particular, note the brightest star which is outside the left edge of the displayed regions.
      Since panel b does not contain noise, its color bar does not match those of the other panels.
  }
\end{figure*}

Fig.~\ref{fig:satghost} illustrates the automated procedure we use for the definition and removal of internal reflections.
Panel a of Fig.~\ref{fig:satghost} shows one of the stars which is contaminated by the internal reflection.
We first divide each image stamp into four quadrants: top right ($0-90$\,degrees), top left ($90-180$\,degrees), lower left ($180-270$\,degrees) and lower right ($270-360$\,degrees).
Radial profiles are created for the flux within each quadrant (see panel b).
We find the radial profile with the lowest $\sigma$-clipped mean value as the reference, as it is the cleanest from reflections.
For the example star of Fig.~\ref{fig:satghost}, the bottom right region ($270-360$\,degrees, colored in violet) is chosen as best and less contaminated part of the star by the reflection.
Afterwards, an azimuthally symmetric 2D image is created from the selected radial profile.

Subtracting this 2D from the stamp image of the star greatly enhances the ghost (see panel d of Fig.~\ref{fig:satghost}).
We then warp the subtracted image to a scale of $1/2$ (so each output pixel covers $2\times2$ input pixels) and run NoiseChisel \citep{gnuastro, akhlaghi19, akhlaghi19b} to detect the diffuse signal.
Panel e shows that the three separate reflections of this star are now clearly detected after the use of NoiseChisel.
The final result of this part of the analysis is shown in panel f, where the detected pixels are converted to the previous resolution and all the detected areas are masked in the original stamp image.
While the internal reflections and wings of fainter stars have been nicely masked the spikes are kept in panel f.
The technical details and implementation steps of finding the spikes and un-masking them are provided in Appendix \ref{ap:spik} and Fig.~\ref{fig:spikeangle}.

After selecting, centering, ranking the brightest stars, and masking the mirror reflections, we proceed to construct the outer part of the PSF.
This process is carried out by normalizing the image stamps (see Sect.~\ref{sec:createcenter}).
It should be noted that during the normalization step, mirror reflections appear mostly in the outer regions of the PSF.
Even after masking, pixels with a high signal-to-noise ratio remain, most of which do not include mirror reflections.
These pixels are then used to determine the normalization value.
Finally, normalization is followed by co-adding the $\numout$ brightest stars (see Fig.~\ref{fig:flowchart}).
It is important to note that this number applies only to the current tile, and the number of bright stars may vary across different tiles.
The resulting radial profile of the outer part of the Coma cluster PSF, derived in the {\flowchartfilter} filter, is illustrated by the red line in Fig.~\ref{fig:radone}.

It may happen that a bright star falls on a large-scale diffuse emission, such as cirrus, galaxy halos, and intra-cluster light (ICL), see \citet{roman20} and \citet{li25}.
If that star is the only bright star used for the outer PSF, these sources of diffuse emission will affect extended PSF reconstruction.
But we usually have more than one star in a very different part of the equatorial sky (another exposure: similar atmosphere conditions, but different RA,DEC).
Since they are not affected by the same type if diffuse emission, the effect on the final coadd of the outer part will be removed (or dramatically decreased).

\begin{figure*}[!t]
  \begin{center}
  \ifdefined\makepdf%
    \tikzsetnextfilename{fig-psf-slope}%
    \input{tex/src/fig-psf-slope.tex}%
  \else
    \includegraphics[width=\linewidth]{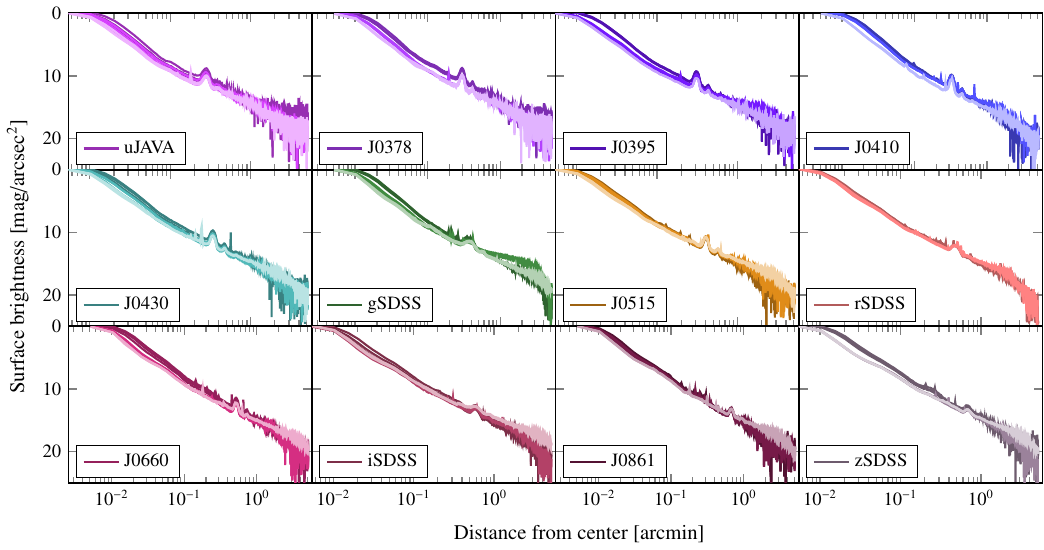}
  \fi

  \end{center}
  \caption{\label{img:radall} Radial profile of three different tiles in different exposures in 12 bands.
    The broad-band PSFs have one peak while the narrow-bands one have two peaks.
    It is due to the ring around star.
    The position of the ring is changed based on the filter, becoming larger from blue to red filters.}
\end{figure*}

\subsection{Constructing the final PSF}
\label{sec:finalpsf}

The united PSF for each exposure is constructed by combining the individually generated central, middle, and outer components, as described earlier.
The approach used to merge these components is another key innovation of this work: instead of applying fixed radii to determine the normalization factor, we adopt a dynamic method based on overlapping pixels between the different parts of the PSF, similar to the technique outlined Sect.~\ref{sec:createcenter}.
Specifically, the middle component is first merged with the central component, followed by the integration of the outer component.
The ``United  PSF'' box in Fig.~\ref{fig:flowchart} is illustrated on the left-hand side of the last row.
In the given example, the united PSF spans a surface brightness range of {$\psfsurfacebrightness$}\,mag\,arcsec$^{-2}$, extending from the center to the outer regions, and covers a radial range of {$\psfsizeouter$} arcminutes.

Ultimately, to obtain the ``Projected PSF'', the 1D radial profile of the united PSF is projected into a two-dimensional circular image using the \texttt{-{}-customtable} feature of Gnuastro’s \textsf{MakeProfiles} \citep[Sec.~8.1 of][]{gnuastrobook}.
This result is displayed on the right-hand side of the last row in Fig.~\ref{fig:flowchart} as ``Projected PSF''.
The projected PSF is necessary in this work because of the relatively small number of stars in the 5 hour interval chosen here: not fully covering the area of the PSF (white regions in the ``united PSF'' of Fig.~\ref{fig:flowchart}).
If there are a sufficient number of bright stars to cover the whole PSF area, the ``Projected PSF'' will not be necessary.

To clarify, the spikes are included in the united PSF, meaning that their pixel values are also considered in projected PSF (for more details, see Appendix \ref{ap:spik}).
Since the spikes are a part of the PSF and their pixel values contribute to the scaling, it is preferable to include them in both the normalization and subtraction steps.
This process enables reconstruction of the PSF, particularly in regions affected by mirror reflections and saturated pixels that have been masked.

\section{Subtracting the PSF}\label{sec:subtract}

To remove the final PSF in each exposure, an extended PSF is first constructed in each exposure.
The scale factor is then determined using the ``United PSF''.
In the next step, the ``Projected PSF'' is scaled for each star and subsequently subtracted.
Before explaining the procedure for determining the scale factor for subtraction, it is important to note that both the background and the foreground objects surrounding each star are masked using the \texttt{ast\-script\--psf\--stamp} tool.
This tool is particularly relevant to non-isolated stars whose outer wings overlap, as it allows other sources to be masked.
In this way, accurate normalization can be performed.

To calculate this scale, we first determine the median value of the sky along with its standard deviation.
The median of the sky’s standard deviation is then multiplied by 3.5 and added to the median sky value, which is defined as the threshold value.
In the next step, for both the united PSF and the star image, all pixels below this threshold are masked (hereafter referred to as the threshold-PSF and threshold-star).
It is important to emphasize that the normalization is not based on radius; rather, only the pixels below this threshold are retained, and the corresponding area varies according to the magnitude of the stars.
To obtain the scale factor from the thresholded images we do not use the raw pixels of the united PSF and star (which can be very different), but the difference of each pixel with its 4-connected neighbors \citep[see fig.~6 of][]{gnuastro} for the united PSF and star (which is more precise, since it is relative not absolute).
This produces four images for each neighbor for the PSF and star (eight images in total).
We then divide the PSF and star image of each neighbor to find the normalization value for that neighbor.
This value is approximately the same in all pixels (which had a high signal-to-noise ratio), so, the sigma-clipped median of all the pixels is used to obtain a single value for that neighbor.
Ultimately, the normalization value for a certain star is found by taking the sigma-clipped median of the four neighbor normalization values.

To subtract the stars in the field, the selected stars in a certain arbitrary magnitude range will be subtracted based on their brightness: the brightest star is first subtracted, followed by the next brightest star and so on.
An example of this PSF subtraction is demonstrated in Fig.~\ref{fig:psfsub}.
NGC 4212 is shown before and after PSF subtraction, also including the scattered light field from the stars (that has been subtracted).

Panel b in Fig.~\ref{fig:psfsub} displays scattered light from stars located at the left edge of the image, partially outside the field of view.
Despite not being fully visible, their light significantly affects the NGC 4212 galaxy.
This highlights the importance of accurately constructing and subtracting the extended PSF, as well as the considerable extent to which stellar light can scatter.

\section{Results}
\label{sec:results}
In this section, we investigate how the PSF changes with filter, time and CCD position.
These parameters are known to affect the shape of the PSF but our automated method on a wide area survey like J-PAS allows us to study the relevant characteristics on a large number of different scenarios.

\subsection{Dependence on filter}
\label{sec:dependonfilter}
J-PLUS employs 12 filters, including both broad- and the narrow-band.
Images of stars have different properties in different filters depending on the optics of the telescope.
In Fig.~\ref{img:radall}, we illustrate how the PSF varies across different filters by presenting the radial profiles of the PSFs obtained from three distinct exposures in three separate tiles.

In particular, we see a secondary peak/ring in the narrow-bands, whereas in the broadband filters, either only one thicker ring is visible or the second ring is much smaller than the first ring (for example, in gSDSS and uJAVA).
This peak is located within a radius of approximately 50 arcseconds in the broad uJAVA filter, whereas in the redder filters, the peaks shift to a radius of 55 to 80 arcseconds.
The larger/shared peak is narrow in the narrow bands but becomes wider in the broad bands.
This is because in the broad bands the wavelength range is wide enough that the two rings, whose diameter changes with wavelength, are washed out and are observed as one single ring but broader ring.
The rings of the light found around bright stars are related to the diffraction pattern and the PSF shape; the diameter is proportional to the wavelength.
These rings, of which there can be one, two, or more, have their origin in the obscuration caused by the secondary mirror and the intermediate baffle situated between the primary and secondary mirrors.
In other telescopes we do not see these rings so significantly, because in JAST80 the secondary mirror is very large in relation to the primary mirror.
The increased size of the secondary mirror is due to our optics being 'faster', resulting in a 'shorter' telescope.
Also, a baffle is installed to prevent direct light from entering the camera.

Another feature is the variation in the shape of the outer PSF wing across filters and tiles.
At the outermost radii of some profiles in Fig.~\ref{img:radall}, such as gSDSS and rSDSS, a drop is observed.
This occurs due to over-subtraction of the sky, meaning that part of the PSF flux has been mistakenly removed as background.
However, this effect is not consistent across all cases, even within the same filter; for example, the light-green profile of gSDSS does not show such a drop.
This issue will be addressed in a future version of the pipeline.

In addition, the number of stars used to construct different parts of the PSF varies across tiles and filters.
Fig.~\ref{img:radall} shows that, although the number of stars differs for each part, these variations do not lead to substantial differences between the resulting PSFs; the only notable difference appears in the outer region, which, as mentioned, is caused by sky over-subtraction.

\begin{figure}[!t]
  \begin{center}
  \ifdefined\makepdf%
    \tikzsetnextfilename{fig-psf-depend-time}%
    \input{tex/src/fig-psf-depend-time.tex}%
  \else
    \includegraphics[width=\linewidth]{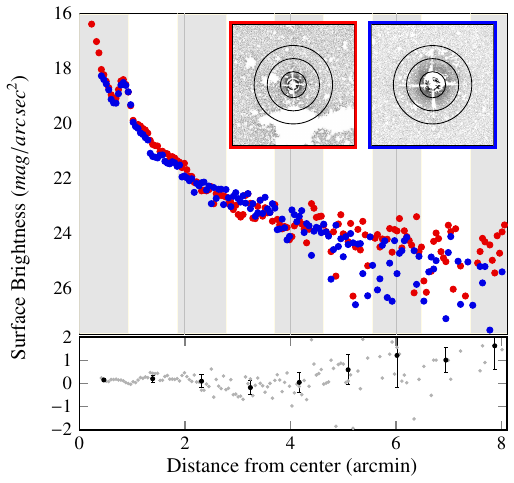}
  \fi

  \end{center}
  \caption{\label{fig:psfdependontime} Effect of time on the PSF: different stars in same position on detector, imaged two hours apart in rSDSS band.
    In both of them extra objects and internal reflection are masked, and the brightest star (blue) is normalized to the faint star (red).
    The size of the black circles are 2, 4 and 6 arcmin.
    The surface brightness radial profile of each star is shown in its own color.
    The surface brightness differences are shown in the lower panel, where the difference between the two stars remains nearly zero within a radius of 4 arcminutes.
    The error bars in this panel were calculated using the standard deviation after applying median absolute deviation (MAD) clipping to the surface brightness differences across every 50 radii.
    The 4 arcmin limit is imposed by the brightness of the fainter star (if it was brighter, we could verify at a larger distance).
    This shows that the PSF does not change significantly in this time interval until this radius.}
\end{figure}

\begin{figure}[!t]
  \begin{center}
  \ifdefined\makepdf%
    \tikzsetnextfilename{fig-psf-depend-pos}%
    \input{tex/src/fig-psf-depend-pos.tex}%
  \else
    \includegraphics[width=\linewidth]{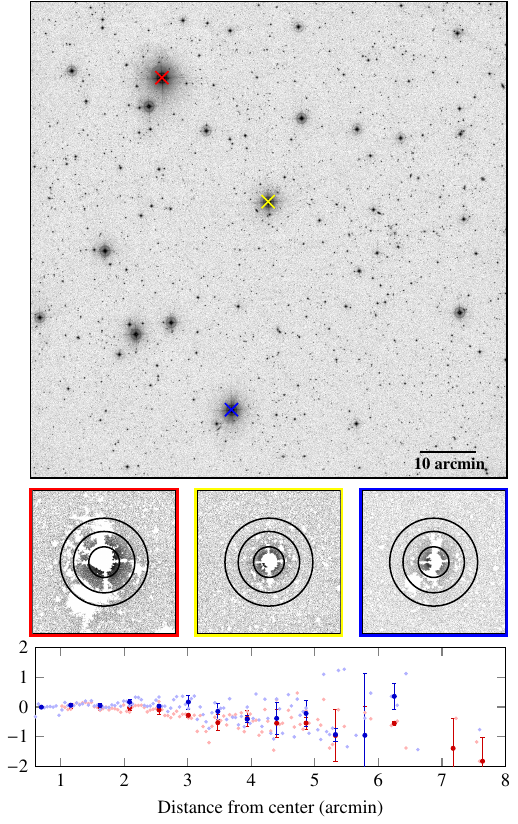}
  \fi

  \end{center}
  \caption{\label{fig:psfdependonposition} Dependence of PSF on position in field of view.
    This tile is chosen because it has many stars which are located in different CCD position.
    The red, yellow and blue stars are chosen to check how PSF will change based on the CCD position.
    In all of them extra object and internal reflection are masked and showed with same color in the middle.
    Black circles depict to 2, 4 and 6 arcminutes radius.
    The yellow star is considered as reference star and the other ones normalize to it and the difference of surface brightness at different radii are shown at the bottom panel.
    Difference at shorter radius than 5 arcminutes is less than 0.5 mag/arcsec$^2$}
\end{figure}

\begin{figure*}[!t]
  \begin{center}
  \ifdefined\makepdf%
    \tikzsetnextfilename{fig-psf-sky-estimate}%
    \input{tex/src/fig-psf-sky-estimate.tex}%
  \else
    \includegraphics[width=\linewidth]{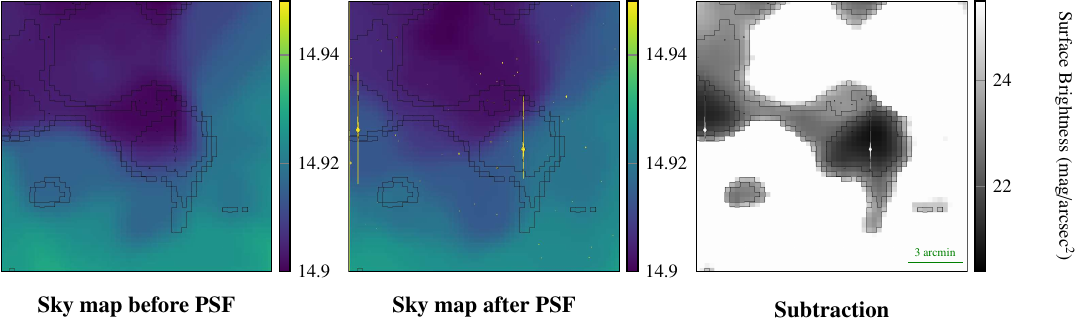}
  \fi

  \end{center}
  \caption{\label{fig:psfskyestimate}
    Effect of PSF subtraction on the estimation of the sky level.
    The left panel shows the sky map from NoiseChisel before PSF subtraction, while the middle one shows it after subtraction.
    The right panel shows the difference between these two panels and all the contours corresponding to the surface brightness of 24, 23, 22\,mag\,arcsec$^{-2}$.}
\end{figure*}

\subsection{Dependence of PSF on time}
\label{sec:dependontime}

To generate the extended PSF, we require bright stars.
However, the availability of stars that are bright enough is restricted within each exposure.
We thus use stars imaged in diverse fields in a certain time range around the exposure to be corrected, to enhance the signal-to-noise ratio.
We assume that the PSF is stable enough to do this, and ascertain whether the PSF undergoes changes across different time frames.
In this section, we analyze the variations in the PSF to acquire a comprehensive understanding of its dynamic behavior over time.

We use two stars, highlighted in red and blue in Fig.~\ref{fig:psfdependontime}, situated at approximately the same CCD position but observed at different times: 2 hours apart in the rSDSS filter.
We generate image stamps of both stars and mask out additional objects and mirror reflections.
We depict the resulting images of the stars (indicated by red and blue) in Fig.~\ref{fig:psfdependontime}, where the black circles illustrate radii of 2, 4, and 6\,arcmin.
After normalization, we show the radial surface brightness profiles of both stars in Fig.~\ref{fig:psfdependontime}, again in red and blue, along with the variation in surface brightness at each radius.

We find that the difference in the PSF between the two stars observed two hours apart, shown in the lower panel of Fig.~\ref{fig:psfdependontime}, is negligible at radii smaller than 4\,arcmin.
At larger radii, the discrepancy is dominated by noise from the fainter star, which may result from sky over-subtraction, as discussed in Sect.~\ref{sec:dependonfilter} regarding Fig.~\ref{img:radall}.
  Further tests will be conducted on this in future versions of this pipeline.
  The error bars in this panel were calculated using the standard deviation after applying median absolute deviation (MAD) clipping to the surface brightness differences across every 50 radii.
This indicates that stars captured at two-hours intervals do not exhibit differences in surface brightness.
The PSF of J-PLUS exposures does not differ much based on time, at least in this test: the conditions for doing observations have successfully removed outliers.
We note that this is a single test and the PSF may not be this stable in other exposures.
In next versions, we will study the expansion of the time interval for selecting stars to derive the outer part of the PSF to be able to incorporate more stars to enhance the contiguity (no blank regions) of the united PSF and its signal-to-noise ratio.

\subsection{Dependence on position}

To generate the PSF, we use stars at various positions across the field of view.
It is well known that reflections of the mirror vary with field of view position, but they are not part of the PSF and as shown in Fig.~\ref{fig:satghost}, they are completely masked in all observed stars.
In this section, we investigate whether any discernible change in the PSF occurs relative to the field of view position in the outer parts of the PSF.

For this investigation, we use the tile displayed in Fig.~\ref{fig:psfdependonposition}, as observed in the rSDSS filter.
This tile was chosen primarily for its bright stars situated at various positions across the field of view.
The three stars marked with colored crosses are used for this test.
The yellow star is used as the reference due to its proximity to the center of the field of view.
We normalized the stars marked in blue and red after masking the mirror reflection and additional objects, to match the yellow star.
The distance of the red object from yellow is 20 arcminutes, while the blue one is 30 arcminutes away from yellow.
We show images of the three stars, before normalization, in the middle panel of Fig.~\ref{fig:psfdependonposition}, where the different black circles illustrate radii of 2, 4, and 6\,arcmin.

The lower panel of Fig.~\ref{fig:psfdependonposition} displays the variation in surface brightness between the blue and red stars compared to the yellow one.
A caveat here is that the reference star, shown in yellow, is not the brightest one, which makes the analysis more challenging.
The blue profile residual lies almost directly above the red one, which is because the blue star is shallower than the red star.
In the future, this will be studied in more detail by constructing dedicated PSFs in different parts of the image from more exposures.
Again, beyond a radius of about 4 arcminutes, the difference is dominated by noise and is not reliable.
But within 4 arcminutes, we see that the outer PSF does not exhibit significant variation with position on the field of view.
This test verifies that our usage of all stars positioned across the various field of view positions to construct the outer PSF.
A crucial aspect of this work is that we entirely masked the mirror reflection in this study.
Without this accurate and robust masking, the PSF would indeed undergo changes with position.

\subsection{Sky estimation and stacks}

The extended PSF of bright stars is a considerable factor in robust sky measurements during the reduction phase of any survey \citep{borlaff22, liu23, watkins24}.
To study this effect we subtracted the sky image provided by \textsf{NoiseChisel} on each exposure and coadded the three exposures.
In one round there was no PSF subtraction and in the second round we subtracted the extended PSF before estimating the sky.

To show the impact of the PSF subtraction, three different exposures for one tile are coadded.
Fig.~\ref{fig:psfskyestimate} highlights the influence of scattered light from stars on the sky map.
The left panel displays the sky map from \textsf{NoiseChisel} before subtraction of the PSF, while the middle panel depicts the map after PSF subtraction.
The right panel corresponds to the difference between the two sky maps: one before and one after PSF subtraction.
The contours in the last panel represent varying levels of surface brightness (24, 23, 22\,mag\,arcsec$^{-2}$).

In the left-hand map, the presence of a bright star is evident, yet after PSF subtraction the sky map appears more uniform.
As illustrated in Fig.~\ref{fig:psfsub}, PSFs outside the field can also contribute, particularly when bright stars are located near the field edges.
Consequently, the potential impact of stars situated beyond the field of view on sky estimation values should not be overlooked.
This observation underscores the importance of PSF subtraction in accurately assessing sky values.

An important point to highlight here is that we are using \textsf{NoiseChisel} is being used to measure the sky.
This improvement will be much more prominent for surveys that use \textsf{SExtractor} \citep{sextractor} in their sky subtraction because \textsf{NoiseChisel} is known to be much more precise in this \citep[see ][]{gnuastro}.

\subsection{Effect on nearby galaxy surface brightness profiles}

\begin{figure}
  \begin{center}
  \ifdefined\makepdf%
    \tikzsetnextfilename{fig-ngc-4212}%
    \input{tex/src/fig-ngc-4212.tex}%
  \else
    \includegraphics[width=\linewidth]{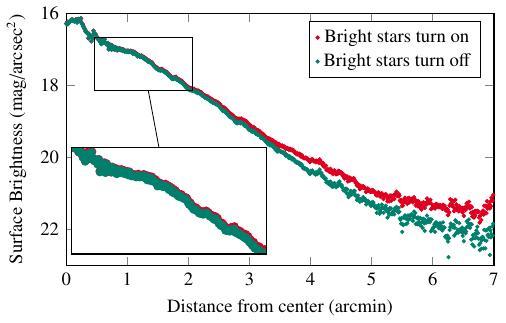}
  \fi

  \end{center}
  \caption{\label{fig:ngc4212} Effect of PSF on the radial profile of NGC 4212 before (red) and after (green) PSF subtraction.
    The 2D effect was shown in Fig.~\ref{fig:psfsub}.
    The PSF affects the whole galaxy from the central to the outer parts (much more significantly on the outer parts).
  }
\end{figure}

In Fig~\ref{fig:ngc4212}, the surface brightness radial profile of the NGC 4212 (which was shown in Fig.~\ref{fig:psfsub}) in a single exposure and in the rSDSS filter is shown before and after PSF subtraction.
The red profile corresponds to the surface brightness of NGC 4212 before PSF subtraction, whereas the red profile shows it after PSF subtraction.
Zooming into the central part of the profile reveals that the PSF of nearby stars also affects the central part (although indeed at a smaller scale).
From the center outward, the difference becomes more significant.

It is important to note that scattered light in a galaxy profile originates not only from the stars but also from their bulges or prominent central regions.
In Fig~\ref{fig:ngc4212}, we considered only the scattered light from the stars and not from the galaxy's center.
This is because our goal here is only to remove the stars, not full scatter light.
Including the scattered light from the bulge is necessary in a dedicated study of nearby galaxies, as it affects all radii at all azimuthal angles \citep{dejong08, sandin14}, whereas the scattered light from stars that are not exactly in the galaxy center will not have this symmetricity.

Until now, galaxies like this (that have a bright star nearby) were usually discarded in most statistical studies during sample selection.
But as shown here, through the robust estimation and subtraction of the extended PSF, we are able to include them in our sample selection.

\section{Discussion and summary}
\label{sec:summaryandconclusion}

The vast majority of surveys tend to focus solely on the central few arcsec of the PSF in their standard data products.
This overlooks the critical need for comprehensive characterization and removal of scattered light (PSF) to much larger distances across the field, particularly stemming from bright sources within the image.
Additionally, each filter exhibits a distinct PSF, adding another layer of complexity to the issue.
Moreover, the diverse colors exhibited by these bright sources introduce a complex, multi-color interplay within the scattered light field, potentially introducing systematic biases in both photometry and color-based measurements of background astrophysical sources.

Large-scale multi-color surveys such as J-PLUS present a unique opportunity with their wide coverage across the sky and numerous filters.
The extended PSF of bright stars is a problem in this survey (as in other wide-field surveys: more significant as they go deeper): affecting 10\% of its total area.
In this paper we demonstrated the unique features of our pipeline to estimate and subtract the extended PSF.
Our pipeline is based on the earlier work of I20 with the following novelties:

\begin{itemize}
\item The extended PSF is estimated and subtracted on a per-exposure level.
\item We do not use radii to normalize stars.
  This happened in three different parts of the process: when coadding the stars to create each part, when uniting the various parts and when normalizing the PSF to each star that was to be subtracted.
  The manually inserted radii were a major issue in J-PLUS because of the diversity of the filters (some broad, some medium and some narrow-band).
\item A robust solution to the problem of internal reflections (ghosts) has been implemented.
\item To determine the central coordinates of saturated stars, we utilized the pixel distribution along the x and y axes within the bleeding region.
\item Due to the limitations caused by stellar emission and absorption lines, as well as the constraints of narrow-band filters, Gaia is not used for sorting bright stars.
  Consequently, a new algorithm has been developed to address this issue without ancillary data.
\item We investigated how the PSF changes from one exposure to another, depending on the CCD position and time.
\item We also examined the effects of PSF on the sky background and the radial profile of NGC 4212.
\item Low-level scripts have been added to Gnuastro \citep[Sect.~10.8 of][]{gnuastrobook} with an extensive documentation and a hands-on tutorial \citep[Sect.~2.3 of][]{gnuastrobook} to simplify the implementation/customization of this method on other surveys.
\item We construct PSF profiles reaching up to {$\psfsurfacebrightness$}\,mag\,arcsec$^{-2}$ in surface brightness and extending up to {$\psfsize$}\,arcmin for each individual J-PLUS DR3 exposure.

\end{itemize}

The J-PLUS PSF subtraction pipeline that was introduced here is written in a generic and reproducible format using Maneage \citep{maneage}.
While it was implemented as a proof of concept in this paper (only on 5 J-PLUS tiles) it is automatic and we plan to apply it on the full J-PLUS in future releases.
Furthermore, there are several improvement which will be implemented in future.
For example the sky subtraction issue mentioned at the end of \ref{sec:dependonfilter}, and optimization in speed and memory usage.

A crucial aspect that warrants detailed examination in a separate study is validation through simulations.
Such simulations must account for several relevant conditions, including lunar distance, zodiacal light, atmospheric fluctuations, airglow, and stray light.
This task becomes particularly complex in the case of J-PLUS, as it employs narrow-, medium-, and broad-band filters, each producing PSFs of varying shapes.
Additionally, internal reflections depend on the CCD position, while diffraction spikes shift accordingly.
Another topic for further study is the dependence on time and detector position in all filters (here we just focused on rSDSS).

The most significant part of the pipeline that is custom-written for J-PLUS is the interface to the database for data access and downloading.
The lower-level components have also been implemented in Gnuastro and easily usable for any instrument.
This will allow its adoption to other large surveys like J-PAS, Euclid Wide Survey or Large Synoptic Survey Telescope (LSST) and other projects like James Webb Space Telescope (JWST), and Analysis of Resolved Remnants of Accreted galaxies as a Key Instrument for Halo Surveys (ARRAKIHS).
Apart from how the data is accessed, how it is named, and which format is used to save it, all the configuration parameters that need to be changed for a different survey are in the \texttt{reproduce/analysis/config} directory.

Through the robust PSF-subtraction that was done here, almost 10\% of the area of J-PLUS will become usable (source near bright stars will not be flagged) and low surface brightness optimized science cases like planetary nebulae, cirrus and the halos of galaxies in the local universe (among others) will become possible with J-PLUS.

The density of brighter stars increases as we go to lower galactic latitudes, causing the loss of many galaxies in those regions of the sky.
Furthermore, with an automated and optimized pipeline that robustly subtracts the PSF (like the one introduced here), it will be possible to go closer to the Galactic plane.
Our findings underscore the significant impact of PSF subtraction on ensure the accurate generation of sky maps and precise determination of object color, thus enhancing the reliability and robustness of large surveys.

\section{Code and data availability}

This project was developed within the reproducible framework of Maneage \citep[Managing data lineage,][latest Maneage commit \texttt{\maneageversion}, dated \maneagedate]{maneage}.
The implementation was carried out on an {\machinearchitecture} architecture with {\machinebyteorder} byte order.
A complete list of the software packages employed, together with their specific versions, is provided in Appendix \ref{appendix:software}.
The source code for the project is publicly available on SoftwareHeritage for longevity as \href{https://archive.softwareheritage.org/swh:1:dir:fc95ed9ad173de6fa64690e3d71ab041c630f32c;origin=https://gitlab.com/Sepideh.Esk/psf-j-plus;visit=swh:1:snp:31bb4550902f7902a0a65d0674106271de776d3e;anchor=swh:1:rev:4860c70d6285a60300e429889e8637e98568a915}{\texttt{swh:1:\-dir:fc95e\-d9ad173de6\-fa64690e\-3d71ab041\-c630f32c}}\footnote{SoftWare Hash IDentifier (SWHID) can be used with resolvers like \texttt{http://n2t.net/} (e.g., \texttt{http://n2t.net/swh:1:...}). Clicking on the SWHIDs will provide more ``context'' for same content.} and all the output data products are available on \href{https://doi.org/10.5281/zenodo.17348653}{Zenodo:17348653}\footnote{\url{https://doi.org/10.5281/zenodo.17348653}}.

\vspace{5mm}

\begin{acknowledgements}
We would like to extend our sincere gratitude to Ignacio Trujillo and the anonymous referee for their invaluable comments and contributions throughout this project.
The authors acknowledge PID2024-162229NB-I00, CNS2023-145339, PID2022-136505NB-I00 and ICTS-MRR-2021-03-CEFCA from the Spanish Ministry of Science and Innovation (MCIN/AEI/10.13039/501100011033).
Co-funded by the European Union (EU) MSCA EDUCADO, GA 101119830.
Views and opinions expressed are however those of the author(s) only and do not necessarily reflect those of the EU. Neither the EU nor the granting authority can be held responsible for them.
Based on observations made with the JAST80 telescope and T80Cam camera for the J-PLUS project at the Observatorio Astrof\'{\i}sico de Javalambre (OAJ), in Teruel, owned, managed, and operated by the Centro de Estudios de F\'{\i}sica del Cosmos de Arag\'on (CEFCA). We acknowledge the OAJ Data Processing and Archiving Unit (UPAD; \citealt{upad}) for reducing the OAJ data used in this work. Funding for the J-PLUS Project has been provided by the Governments of Spain and Arag\'on through the Fondo de Inversiones de Teruel; the Aragonese Government through the Research Groups E96, E103, E16\_17R, E16\_20R, and E16\_23R; the MCIN/AEI y FEDER, Una manera de hacer Europa) with grants PID2021-124918NB-C41, PID2021-124918NB-C42, PID2021-124918NA-C43, and PID2021-124918NB-C44; the Spanish Ministry of Science, Innovation and Universities (MCIU/AEI/FEDER, UE) with grants PGC2018-097585-B-C21 and PGC2018-097585-B-C22; the Spanish Ministry of Economy and Competitiveness (MINECO) under AYA2015-66211-C2-1-P, AYA2015-66211-C2-2, AYA2012-30789, and ICTS-2009-14; and European FEDER funding (FCDD10-4E-867, FCDD13-4E-2685). The Brazilian agencies FINEP, FAPESP, and the National Observatory of Brazil have also contributed to this Project.
\end{acknowledgements}

\vspace{5mm}

\bibliographystyle{aa}
\bibliography{paper}

\begin{appendix}
\section{J-PLUS database}
\label{ap:query}

To get the relevant metadata of all J-PLUS exposures, we used the ADQL query below, which can be submitted in the J-PLUS webpage\footnote{\url{https://archive.cefca.es/catalogues/jplus-dr3/tap_async.html}} after login.

\begin{verbatim}
SELECT tile.ref_tile_id, rc.tile_id,
       rc_id, filter.name as filter_name,
       rc.RA,rc.DEC, rc.OBSERVATION_DATE,
       rc.FWHM_MEAN, rc.EXPTIME, rc.AIRMASS,
       tile.name as tile_name, tile.zpt
FROM jplus.ReducedIndividualFrame AS rc
     JOIN jplus.Filter AS filter
          ON filter.filter_id = rc.filter_id
     JOIN jplus.TileImage AS tile
          ON tile.tile_id = rc.tile_id
\end{verbatim}

The output of the above query identifies each exposure with an ``RC-ID''.
The required exposures can be downloaded from a generic link, which must be modified by replacing the RC-ID with the corresponding exposure number.
For example, to obtain the image with an RC-ID of 1040978, replace RC-ID with 1040978 in the link provided\footnote{URL: \url{http://archive.cefca.es/catalogues/vo/siap/jplus-dr3/reduced/get_fits?id=RC-ID}}.

\section{Identifying Stars Spikes}
\label{ap:spik}

In Fig.~\ref{fig:satghost} we demonstrated the process of masking internal reflections (ghosts).
However, the spikes produced by bright stars were also masked in the process.
These spikes come from diffraction and are formed by the secondary mirror and baffle holders.
Thus, the spikes are a component of the PSF.
However, the reflection from the mirror does spread the light and thus is not considered part of the PSF.
Therefore, spikes should not be masked out before deriving the PSF, whereas the mirror's reflection should be masked.

\begin{figure}[!t]
  \begin{center}
  \ifdefined\makepdf%
    \tikzsetnextfilename{fig-spike-angle}%
    \input{tex/src/fig-spike-angle.tex}%
  \else
    \includegraphics[width=\linewidth]{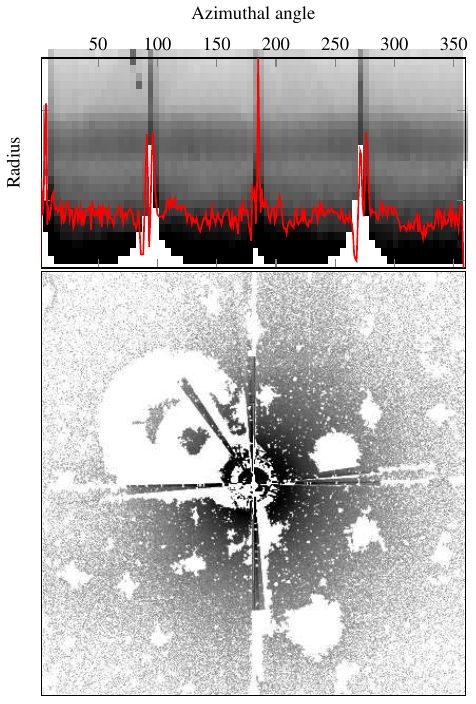}
  \fi

  \end{center}
  \captionof{figure}{\label{fig:spikeangle} Un-masking the spikes in the image of bright stars is done using polar plots. The top panel shows the output of polar plot. The x-axis shows azimuthal angle and y-axis show radius. The white pixels were originally bleeding pixels. The red profile shows the collapsed value at each azimuthal angle. Based on this plot the angle of the spike are extracted and removed from the masked pixels shows in Fig.~\ref{fig:satghost}(e).}
\end{figure}

When a structure depends on distance and azimuthal angle from a central point, such as the diffraction spikes of stars, projecting the pixels to polar coordinates significantly simplifies the analysis.
For this project, a new option for generating polar plots has been added to Gnuastro \citep[see][]{Eskandarlou24}.
To restore the spikes, we create a polar plot of each star stamp after masking the internal reflection.
We show in the top panel of Fig.~\ref{fig:spikeangle} the polar plot of the stamp image in Fig.~\ref{fig:satghost}(f).
We then collapse the polar plot in the radius axis with a $\sigma$-clipped mean, which is shown as the red curve overlaid on the top panel of Fig.~\ref{fig:spikeangle}.
At the angles where spikes exist, a strong peak is present.
Based on the discovered angles, the pixels of the spikes are removed from the mask of Fig.~\ref{fig:satghost}(e).
The lower part of Fig.~\ref{fig:spikeangle} shows how the internal reflection is masked but the spikes are kept.

However, we found that the spikes from bright stars depend on the CCD position and should be incorporated during the PSF subtraction step to accurately derive the final PSF.
Since the spikes are a part of the PSF, their pixel values are included throughout the process to improve normalization.
Nevertheless, due to the reason mentioned above, they are currently not be subtracted in the final stage of PSF subtraction.
Their full implementation in the subtraction process will be addressed in future studies.

\section{Software acknowledgement}
\label{appendix:software}
\input{tex/build/macros/dependencies.tex}

\end{appendix}

\end{document}

%% file: tex/build/macros/project.tex
\input{tex/build/macros/initialize.tex}

\input{tex/build/macros/psf-create.tex}

\input{tex/build/macros/psf-subtract.tex}

\input{tex/build/macros/psf-slope.tex}
\input{tex/build/macros/fig-flowchart-psf-create.tex}
\input{tex/build/macros/fig-radial-profile.tex}
\input{tex/build/macros/fig-sat-obj-coord.tex}
\input{tex/build/macros/fig-sort-bright-sat-star.tex}
\input{tex/build/macros/fig-sat-ghost.tex}
\input{tex/build/macros/fig-spike-angle.tex}
\input{tex/build/macros/fig-histogram-sb.tex}
\input{tex/build/macros/fig-radial-profile-all-filter.tex}
\input{tex/build/macros/fig-psf-subtract.tex}
\input{tex/build/macros/fig-psf-depend-pos.tex}
\input{tex/build/macros/fig-psf-depend-time.tex}
\input{tex/build/macros/fig-psf-sky-estimate.tex}
\input{tex/build/macros/fig-ngc-4212.tex}
\input{tex/build/macros/verify.tex}
\input{tex/build/macros/hardware-parameters.tex}

%% file: tex/build/macros/initialize.tex
\newcommand{\projecttitle}{J-PLUS: Turning Off the Bright Stars}
\newcommand{\projectversion}{4860c70}
\newcommand{\projectgitrepo}{https://gitlab.com/Sepideh.Esk/psf-j-plus}
\newcommand{\projectcopyrightowner}{Sepideh Eskandarlou <sepideh.eskandarlou@gmail.com>}
\newcommand{\maneagedate}{12 May 2025}
\newcommand{\maneageversion}{a575ef8}
\newcommand{\satareainpix}{50}
\newcommand{\outmultipleradius}{2}

\newcommand{\sortsatstarwidth}{250}

%% file: tex/build/macros/psf-create.tex
\newcommand{\psfnumbercenter}{100}
\newcommand{\numout}{22}

\newcommand{\psfnumbermiddle}{50}
\newcommand{\psfsizecenterpaper}{0.46}
\newcommand{\psfsizemiddlepaper}{0.92}
\newcommand{\psfsizeouter}{9.17}
\newcommand{\normsigmaout}{5}
\newcommand{\psfdeltahour}{5}

\newcommand{\flowcharttilerefid}{85369}
\newcommand{\flowchartfilter}{rSDSS}
\newcommand{\numtilespaper}{935}

%% file: tex/build/macros/psf-subtract.tex
\newcommand{\psfsurfacebrightness}{15}
\newcommand{\psfsize}{4}

%% file: tex/build/macros/hardware-parameters.tex
\newcommand{\machinearchitecture}{x86\_64}
\newcommand{\machinebyteorder}{Little Endian}

%% file: tex/src/preamble-maneage.tex
\ifdefined\highlightnew

\else

\fi

\ifdefined\highlightnotes
\newcommand{\tonote}[1]{\textcolor{red!60!black}{[#1]}}
\else
\newcommand{\tonote}[1]{{}}
\fi

%% file: tex/src/preamble-project.tex
\usepackage{graphicx}

\usepackage{fixltx2e}

\usepackage{xcolor}
\color{black} 
\definecolor{DarkBlue}{RGB}{0,0,90}

\usepackage{setspace, caption}
\usepackage[
  colorlinks,
  urlcolor=blue,
  citecolor=blue,
  linkcolor=blue,
  linktocpage]{hyperref}

\hypersetup{
    pdftitle={\projecttitle},
    pdfauthor={\projectcopyrightowner},
    pdfsubject={\projectgitrepo{} (commit \projectversion)},
    pdfkeywords={Reproducible research, Maneage, ADD YOUR OWN}
}

\input{tex/src/preamble-pgfplots.tex}

\usepackage{txfonts}

%% file: tex/src/preamble-pgfplots.tex
\usepackage{tikz}
\usetikzlibrary{external}
\tikzsetexternalprefix{tex/tikz/}

\usepackage{pgfplots}
\pgfplotsset{compat=newest}
\usepgfplotslibrary{groupplots}
\usetikzlibrary{spy}
\pgfplotsset{
  axis line style={thick},
  tick style={semithick},
  tick label style = {font=\footnotesize},
  every axis label = {font=\footnotesize},
  legend style = {font=\footnotesize},
  label style = {font=\footnotesize}
  }

\definecolor{ujavacolor}{rgb}{0.847,0.255,1}
\definecolor{j0378color}{rgb}{0.702,0.255,1}
\definecolor{j0395color}{rgb}{0.459,0.098,1}
\definecolor{j0410color}{rgb}{0.318,0.318,1}
\definecolor{j0430color}{rgb}{0.318,0.725,0.725}
\definecolor{gsdsscolor}{rgb}{0.255,0.549,0.255}
\definecolor{j0515color}{rgb}{0.878,0.549,0.098}
\definecolor{rsdsscolor}{rgb}{1,0.51,0.51}
\definecolor{j0660color}{rgb}{0.835,0.18,0.51}
\definecolor{isdsscolor}{rgb}{0.702,0.255,0.404}
\definecolor{j0861color}{rgb}{0.463,0.102,0.282}
\definecolor{zsdsscolor}{rgb}{0.608,0.51,0.608}

\definecolor{mediumvioletred}{rgb}{0.78,0.08,0.52}
\definecolor{deeppink}{rgb}{1.00,0.08,0.58}
\definecolor{palevioletred}{rgb}{0.86,0.44,0.58}
\definecolor{hotpink}{rgb}{1.00,0.41,0.71}
\definecolor{lightpink}{rgb}{1.00,0.71,0.76}
\definecolor{pink}{rgb}{1.00,0.75,0.80}
\definecolor{darkred}{rgb}{0.55,0.00,0.00}
\definecolor{red}{rgb}{1.00,0.00,0.00}
\definecolor{firebrick}{rgb}{0.70,0.13,0.13}
\definecolor{crimson}{rgb}{0.86,0.08,0.24}
\definecolor{indianred}{rgb}{0.80,0.36,0.36}
\definecolor{lightcoral}{rgb}{0.94,0.50,0.50}
\definecolor{salmon}{rgb}{0.98,0.50,0.45}
\definecolor{darksalmon}{rgb}{0.91,0.59,0.48}
\definecolor{lightsalmon}{rgb}{1.00,0.63,0.48}
\definecolor{orangered}{rgb}{1.00,0.27,0.00}
\definecolor{tomato}{rgb}{1.00,0.39,0.28}
\definecolor{darkorange}{rgb}{1.00,0.55,0.00}
\definecolor{coral}{rgb}{1.00,0.50,0.31}
\definecolor{orange}{rgb}{1.00,0.65,0.00}
\definecolor{darkkhaki}{rgb}{0.74,0.72,0.42}
\definecolor{gold}{rgb}{1.00,0.84,0.00}
\definecolor{khaki}{rgb}{0.94,0.90,0.55}
\definecolor{peachpuff}{rgb}{1.00,0.85,0.73}
\definecolor{yellow}{rgb}{1.00,1.00,0.00}
\definecolor{palegoldenrod}{rgb}{0.93,0.91,0.67}
\definecolor{moccasin}{rgb}{1.00,0.89,0.71}
\definecolor{papayawhip}{rgb}{1.00,0.94,0.84}
\definecolor{lightgoldenrodyellow}{rgb}{0.98,0.98,0.82}
\definecolor{lemonchiffon}{rgb}{1.00,0.98,0.80}
\definecolor{lightyellow}{rgb}{1.00,1.00,0.88}
\definecolor{maroon}{rgb}{0.50,0.00,0.00}
\definecolor{brown}{rgb}{0.65,0.16,0.16}
\definecolor{saddlebrown}{rgb}{0.55,0.27,0.07}
\definecolor{sienna}{rgb}{0.63,0.32,0.18}
\definecolor{chocolate}{rgb}{0.82,0.41,0.12}
\definecolor{darkgoldenrod}{rgb}{0.72,0.53,0.04}
\definecolor{peru}{rgb}{0.80,0.52,0.25}
\definecolor{rosybrown}{rgb}{0.74,0.56,0.56}
\definecolor{goldenrod}{rgb}{0.85,0.65,0.13}
\definecolor{sandybrown}{rgb}{0.96,0.64,0.38}
\definecolor{tan}{rgb}{0.82,0.71,0.55}
\definecolor{burlywood}{rgb}{0.87,0.72,0.53}
\definecolor{wheat}{rgb}{0.96,0.87,0.70}
\definecolor{navajowhite}{rgb}{1.00,0.87,0.68}
\definecolor{bisque}{rgb}{1.00,0.89,0.77}
\definecolor{blanchedalmond}{rgb}{1.00,0.92,0.80}
\definecolor{cornsilk}{rgb}{1.00,0.97,0.86}
\definecolor{darkgreen}{rgb}{0.00,0.39,0.00}
\definecolor{green}{rgb}{0.00,0.50,0.00}
\definecolor{darkolivegreen}{rgb}{0.33,0.42,0.18}
\definecolor{forestgreen}{rgb}{0.13,0.55,0.13}
\definecolor{seagreen}{rgb}{0.18,0.55,0.34}
\definecolor{olive}{rgb}{0.50,0.50,0.00}
\definecolor{olivedrab}{rgb}{0.42,0.56,0.14}
\definecolor{mediumseagreen}{rgb}{0.24,0.70,0.44}
\definecolor{limegreen}{rgb}{0.20,0.80,0.20}
\definecolor{lime}{rgb}{0.00,1.00,0.00}
\definecolor{springgreen}{rgb}{0.00,1.00,0.50}
\definecolor{mediumspringgreen}{rgb}{0.00,0.98,0.60}
\definecolor{darkseagreen}{rgb}{0.56,0.74,0.56}
\definecolor{mediumaquamarine}{rgb}{0.40,0.80,0.67}
\definecolor{yellowgreen}{rgb}{0.60,0.80,0.20}
\definecolor{lawngreen}{rgb}{0.49,0.99,0.00}
\definecolor{chartreuse}{rgb}{0.50,1.00,0.00}
\definecolor{lightgreen}{rgb}{0.56,0.93,0.56}
\definecolor{greenyellow}{rgb}{0.68,1.00,0.18}
\definecolor{palegreen}{rgb}{0.60,0.98,0.60}
\definecolor{teal}{rgb}{0.00,0.50,0.50}
\definecolor{darkcyan}{rgb}{0.00,0.55,0.55}
\definecolor{lightseagreen}{rgb}{0.13,0.70,0.67}
\definecolor{cadetblue}{rgb}{0.37,0.62,0.63}
\definecolor{darkturquoise}{rgb}{0.00,0.81,0.82}
\definecolor{mediumturquoise}{rgb}{0.28,0.82,0.80}
\definecolor{turquoise}{rgb}{0.25,0.88,0.82}
\definecolor{aqua}{rgb}{0.00,1.00,1.00}
\definecolor{cyan}{rgb}{0.00,1.00,1.00}
\definecolor{aquamarine}{rgb}{0.50,1.00,0.83}
\definecolor{paleturquoise}{rgb}{0.69,0.93,0.93}
\definecolor{lightcyan}{rgb}{0.88,1.00,1.00}
\definecolor{midnightblue}{rgb}{0.10,0.10,0.44}
\definecolor{navy}{rgb}{0.00,0.00,0.50}
\definecolor{darkblue}{rgb}{0.00,0.00,0.55}
\definecolor{mediumblue}{rgb}{0.00,0.00,0.80}
\definecolor{blue}{rgb}{0.00,0.00,1.00}
\definecolor{royalblue}{rgb}{0.25,0.41,0.88}
\definecolor{steelblue}{rgb}{0.27,0.51,0.71}
\definecolor{dodgerblue}{rgb}{0.12,0.56,1.00}
\definecolor{deepskyblue}{rgb}{0.00,0.75,1.00}
\definecolor{cornflowerblue}{rgb}{0.39,0.58,0.93}
\definecolor{skyblue}{rgb}{0.53,0.81,0.92}
\definecolor{lightskyblue}{rgb}{0.53,0.81,0.98}
\definecolor{lightsteelblue}{rgb}{0.69,0.77,0.87}
\definecolor{lightblue}{rgb}{0.68,0.85,0.90}
\definecolor{powderblue}{rgb}{0.69,0.88,0.90}
\definecolor{indigo}{rgb}{0.29,0.00,0.51}
\definecolor{purple}{rgb}{0.50,0.00,0.50}
\definecolor{darkmagenta}{rgb}{0.55,0.00,0.55}
\definecolor{darkviolet}{rgb}{0.58,0.00,0.83}
\definecolor{darkslateblue}{rgb}{0.28,0.24,0.55}
\definecolor{blueviolet}{rgb}{0.54,0.17,0.89}
\definecolor{darkorchid}{rgb}{0.60,0.20,0.80}
\definecolor{fuchsia}{rgb}{1.00,0.00,1.00}
\definecolor{magenta}{rgb}{1.00,0.00,1.00}
\definecolor{slateblue}{rgb}{0.42,0.35,0.80}
\definecolor{mediumslateblue}{rgb}{0.48,0.41,0.93}
\definecolor{mediumorchid}{rgb}{0.73,0.33,0.83}
\definecolor{mediumpurple}{rgb}{0.58,0.44,0.86}
\definecolor{orchid}{rgb}{0.85,0.44,0.84}
\definecolor{violet}{rgb}{0.93,0.51,0.93}
\definecolor{plum}{rgb}{0.87,0.63,0.87}
\definecolor{thistle}{rgb}{0.85,0.75,0.85}
\definecolor{lavender}{rgb}{0.90,0.90,0.98}
\definecolor{mistyrose}{rgb}{1.00,0.89,0.88}
\definecolor{antiquewhite}{rgb}{0.98,0.92,0.84}
\definecolor{linen}{rgb}{0.98,0.94,0.90}
\definecolor{beige}{rgb}{0.96,0.96,0.86}
\definecolor{whitesmoke}{rgb}{0.96,0.96,0.96}
\definecolor{lavenderblush}{rgb}{1.00,0.94,0.96}
\definecolor{oldlace}{rgb}{0.99,0.96,0.90}
\definecolor{aliceblue}{rgb}{0.94,0.97,1.00}
\definecolor{seashell}{rgb}{1.00,0.96,0.93}
\definecolor{ghostwhite}{rgb}{0.97,0.97,1.00}
\definecolor{honeydew}{rgb}{0.94,1.00,0.94}
\definecolor{floralwhite}{rgb}{1.00,0.98,0.94}
\definecolor{azure}{rgb}{0.94,1.00,1.00}
\definecolor{mintcream}{rgb}{0.96,1.00,0.98}
\definecolor{snow}{rgb}{1.00,0.98,0.98}
\definecolor{ivory}{rgb}{1.00,1.00,0.94}
\definecolor{white}{rgb}{1.00,1.00,1.00}
\definecolor{black}{rgb}{0.00,0.00,0.00}
\definecolor{darkslategray}{rgb}{0.18,0.31,0.31}
\definecolor{dimgray}{rgb}{0.41,0.41,0.41}
\definecolor{slategray}{rgb}{0.44,0.50,0.56}
\definecolor{gray}{rgb}{0.50,0.50,0.50}
\definecolor{lightslategray}{rgb}{0.47,0.53,0.60}
\definecolor{darkgray}{rgb}{0.66,0.66,0.66}
\definecolor{silver}{rgb}{0.75,0.75,0.75}
\definecolor{lightgray}{rgb}{0.83,0.83,0.83}
\definecolor{gainsboro}{rgb}{0.86,0.86,0.86}

%% file: tex/src/fig-flow-chart.tex
\newcommand{\flowchartexps}{}
\newcommand{\flowchartexpsfiveh}{}
\newcommand{\flowchartmarkexpsfiveh}{}
\newcommand{\flowchartstampouter}{}
\newcommand{\flowchartpsfouter}{}
\newcommand{\flowchartexpsingle}{}
\newcommand{\flowchartstampmid}{}
\newcommand{\flowchartpsfmiddle}{}
\newcommand{\flowchartmarkexpsingle}{}
\newcommand{\flowchartstampcen}{}
\newcommand{\flowchartmarkstackmiddle}{}
\newcommand{\flowchartpsfcenter}{}
\newcommand{\flowchartmarkstackcenter}{}
\newcommand{\flowchartfinalpsf}{}

\newcommand{\addimg}[4]{
  \node[anchor=south] (img) at (#1\linewidth,#2\linewidth)
       {\includegraphics[width=#3\linewidth]
         {tex/build/figures/#4.pdf}};
}

\newcommand{\addarow}[4]{
    \draw[->, thick, line width=1.5pt]
       (#1\linewidth,#2\linewidth)
    -- (#3\linewidth,#4\linewidth);
}

\newcommand{\addredline}[5]{
    \draw[thick, #1, line width=0.2pt]
         (#2\linewidth,#3\linewidth)
      -- (#4\linewidth,#5\linewidth);
}

\newcommand{\addarowcurve}[8]{
  \draw[->, thick, rounded corners, line width=1.5pt]
       (#1\linewidth,#2\linewidth)
       -- (#3\linewidth,#4\linewidth)
       |- (#5\linewidth,#6\linewidth)
       |- (#7\linewidth,#8\linewidth);
}

\newcommand{\addradialprofile}[5]{
    \addplot [#1!#2!#3,line width=3pt] table {tex/build/figures/flow-chart/#4.txt};
    \addlegendentry{#5}
}

\newcommand{\addlabel}[6]{
  \node[anchor=#1, rotate=#2] at (#3\linewidth,#4\linewidth)
       {\textcolor{#5}{#6}};
}


\begin{tikzpicture}


\ifdefined\flowchartexpsingle
  \node [at={(0.15\linewidth,-0.565\linewidth)},
         rectangle,
         very thick,
         text centered,
         rounded corners,
         minimum width=0.48\linewidth,
         minimum height=0.4\linewidth,
         draw=black!50!black!50,
         fill=black!10!gray!10!white] {};
\fi

\ifdefined\flowchartexpsfiveh
  \node [at={(0.648\linewidth,-0.565\linewidth)},
         rectangle,
         very thick,
         text centered,
         rounded corners,
         minimum width=0.49\linewidth,
         minimum height=0.4\linewidth,
         draw=black!50!black!50,
         fill=black!10!gray!10!white] {};
\fi

\ifdefined\flowchartstampcen
  \node [at={(0.11\linewidth,-0.96\linewidth)},
         rectangle,
         very thick,
         text centered,
         rounded corners,
         minimum width=0.27\linewidth,
         minimum height=0.25\linewidth,
         draw=black!50!black!50,
         fill=black!10!gray!10!white] {};
\fi

\ifdefined\flowchartstampmid
  \node [at={(0.395\linewidth,-0.96\linewidth)},
         rectangle,
         very thick,
         text centered,
         rounded corners,
         minimum width=0.27\linewidth,
         minimum height=0.25\linewidth,
         draw=black!50!black!50,
         fill=black!10!gray!10!white] {};
\fi

\ifdefined\flowchartstampouter
  \node [at={(0.69\linewidth,-0.96\linewidth)},
         rectangle,
         very thick,
         text centered,
         rounded corners,
         minimum width=0.29\linewidth,
         minimum height=0.25\linewidth,
         draw=black!50!black!50,
         fill=black!10!gray!10!white] {};
\fi

  \addlabel{west}{0}{-0.1}{0.000}{black}{ }
  \addlabel{east}{0}{0.92}{-1.75}{black}{ }

  \addimg{0.41}{0}{0.3}{flow-chart/stack}
  \addlabel{north west}{0}{0.35}{0.31}{black}{Stack}

\ifdefined\flowchartexps
  \addarow{0.41}{0.014}{0.41}{-0.03}
  \addarowcurve{0.41}{0.014}{0.41}{-0.001}{0.16}{-0.001}{0.16}{-0.03}
  \addarowcurve{0.41}{0.014}{0.41}{-0.001}{0.66}{-0.001}{0.66}{-0.03}
\fi

\ifdefined\flowchartexpsingle
\addarowcurve{0.41}{-0.315}{0.41}{-0.325}{0.16}{-0.325}{0.16}{-0.36}
\fi
\ifdefined\flowchartexpsfiveh
  \addarowcurve{0.41}{-0.315}{0.41}{-0.325}{0.66}{-0.325}{0.66}{-0.36}
\fi

\ifdefined\flowchartstampcen
  \addarow{0.16}{-0.768}{0.16}{-0.83}
\fi
\ifdefined\flowchartstampmid
  \addarowcurve{0.16}{-0.768}{0.16}{-0.78}{0.41}{-0.78}{0.41}{-0.83}
\fi
\ifdefined\flowchartstampouter
  \addarow{0.66}{-0.768}{0.66}{-0.83}
\fi

\ifdefined\flowchartpsfcenter
  \addarow{0.15}{-1.085}{0.15}{-1.170}
\fi
\ifdefined\flowchartpsfmiddle
  \addarow{0.41}{-1.085}{0.41}{-1.150}
\fi
\ifdefined\flowchartpsfouter
  \addarow{0.69}{-1.085}{0.69}{-1.135}
\fi

\ifdefined\flowchartfinalpsf
\addarowcurve{0.15}{-1.373}{0.15}{-1.43}{0.3}{-1.43}{0.3}{-1.46}
\addarowcurve{0.41}{-1.388}{0.41}{-1.43}{0.3}{-1.43}{0.3}{-1.46}
\addarowcurve{0.69}{-1.4}{0.69}{-1.43}{0.3}{-1.43}{0.3}{-1.46}
\addarow{0.39}{-1.6}{0.455}{-1.6}
\fi

\ifdefined\flowchartexps
  \addimg{0.06}{-0.327}{0.28}{flow-chart/exp-1}
  \addimg{0.41}{-0.327}{0.28}{flow-chart/exp-2}
  \addimg{0.76}{-0.327}{0.28}{flow-chart/exp-3}

  \addlabel{north west}{0}{-0.01}{-0.04}{black}{Exp-1}
  \addlabel{north west}{0}{0.340}{-0.04}{black}{Exp-2}
  \addlabel{north west}{0}{0.690}{-0.04}{black}{Exp-3}
\fi

\ifdefined\flowchartexpsfiveh

  \ifdefined\flowchartmarkexpsfiveh
    \addimg{0.455}{-0.5}{0.08}{flow-chart/outer-15-star}
  \else
    \addimg{0.455}{-0.5}{0.08}{flow-chart/out-1}
  \fi

  \ifdefined\flowchartmarkexpsfiveh
    \addimg{0.55}{-0.5}{0.08}{flow-chart/outer-14-star}
  \else
    \addimg{0.55}{-0.5}{0.08}{flow-chart/out-2}
  \fi

  \ifdefined\flowchartmarkexpsfiveh
    \addimg{0.645}{-0.5}{0.08}{flow-chart/outer-16-star}
  \else
    \addimg{0.645}{-0.5}{0.08}{flow-chart/out-3}
  \fi

  \ifdefined\flowchartmarkexpsfiveh
    \addimg{0.743}{-0.5}{0.08}{flow-chart/outer-17-star}
  \else
    \addimg{0.743}{-0.5}{0.08}{flow-chart/out-4}
  \fi

  \ifdefined\flowchartmarkexpsfiveh
    \addimg{0.843}{-0.5}{0.08}{flow-chart/outer-13-star}
  \else
    \addimg{0.843}{-0.59}{0.08}{flow-chart/out-5}
  \fi
  \ifdefined\flowchartmarkexpsfiveh
    \addimg{0.455}{-0.59}{0.08}{flow-chart/outer-19-star}
  \else
    \addimg{0.455}{-0.59}{0.08}{flow-chart/out-6}
  \fi

  \ifdefined\flowchartmarkexpsfiveh
    \addimg{0.55}{-0.59}{0.08}{flow-chart/outer-3-star}
  \else
    \addimg{0.55}{-0.59}{0.08}{flow-chart/out-7}
  \fi

  \ifdefined\flowchartmarkexpsfiveh
    \addimg{0.645}{-0.59}{0.08}{flow-chart/outer-12-star}
  \else
    \addimg{0.645}{-0.59}{0.08}{flow-chart/out-8}
  \fi

  \ifdefined\flowchartmarkexpsfiveh
    \addimg{0.743}{-0.59}{0.08}{flow-chart/outer-1-star}
  \else
    \addimg{0.743}{-0.59}{0.08}{flow-chart/out-9}
  \fi

  \ifdefined\flowchartmarkexpsfiveh
    \addimg{0.843}{-0.59}{0.08}{flow-chart/outer-2-star}
  \else
    \addimg{0.843}{-0.59}{0.08}{flow-chart/out-10}
  \fi
  \ifdefined\flowchartmarkexpsfiveh
    \addimg{0.455}{-0.68}{0.08}{flow-chart/outer-10-star}
  \else
    \addimg{0.455}{-0.68}{0.08}{flow-chart/out-11}
  \fi

  \ifdefined\flowchartmarkexpsfiveh
    \addimg{0.55}{-0.68}{0.08}{flow-chart/outer-9-star}
  \else
    \addimg{0.55}{-0.68}{0.08}{flow-chart/out-12}
  \fi

  \ifdefined\flowchartmarkexpsfiveh
    \addimg{0.645}{-0.68}{0.08}{flow-chart/outer-11-star}
  \else
    \addimg{0.645}{-0.68}{0.08}{flow-chart/out-13}
  \fi

  \ifdefined\flowchartmarkexpsfiveh
    \addimg{0.743}{-0.68}{0.08}{flow-chart/outer-22-star}
  \else
    \addimg{0.743}{-0.68}{0.08}{flow-chart/out-14}
  \fi

  \ifdefined\flowchartmarkexpsfiveh
    \addimg{0.843}{-0.68}{0.08}{flow-chart/outer-22-star}
  \else
    \addimg{0.843}{-0.68}{0.08}{flow-chart/out-15}
  \fi
  \ifdefined\flowchartmarkexpsfiveh
    \addimg{0.455}{-0.77}{0.08}{flow-chart/outer-22-star}
  \else
    \addimg{0.455}{-0.77}{0.08}{flow-chart/out-16}
  \fi

  \ifdefined\flowchartmarkexpsfiveh
    \addimg{0.55}{-0.77}{0.08}{flow-chart/outer-4-star}
  \else
    \addimg{0.55}{-0.77}{0.08}{flow-chart/out-17}
  \fi

  \ifdefined\flowchartmarkexpsfiveh
    \addimg{0.645}{-0.77}{0.08}{flow-chart/outer-5-star}
  \else
    \addimg{0.645}{-0.77}{0.08}{flow-chart/out-18}
  \fi

  \ifdefined\flowchartmarkexpsfiveh
    \addimg{0.743}{-0.77}{0.08}{flow-chart/outer-6-star}
  \else
    \addimg{0.743}{-0.77}{0.08}{flow-chart/out-19}
  \fi

  \addlabel{north west}{0}{0.45}{-0.355}{black}{All exposures within $5$h}
\fi

\ifdefined\flowchartexpsingle
  \ifdefined\flowchartmarkexpsingle
    \addimg{0.15}{-0.768}{0.35}{flow-chart/midcen-mark}
  \else
    \addimg{0.12}{-0.82}{0.35}{flow-chart/exp-3}
  \fi

  \addlabel{north west}{0}{0.02}{-0.355}{black}{Single exposure}
\fi

\ifdefined\flowchartstampouter
 \addimg{0.60}{-0.94}{0.08}{flow-chart/outer-1}
 \addimg{0.69}{-0.94}{0.08}{flow-chart/outer-2}
 \addimg{0.78}{-0.94}{0.08}{flow-chart/outer-3}
 \addimg{0.60}{-1.03}{0.08}{flow-chart/outer-4}
 \addimg{0.69}{-1.03}{0.08}{flow-chart/outer-5}
 \addimg{0.78}{-1.03}{0.08}{flow-chart/outer-6}

 \addlabel{north west}{0}{0.58}{-1.03}{black}{\tiny Stamps of outer}
\fi

\ifdefined\flowchartstampmid
 \addimg{0.300}{-0.91}{0.05}{flow-chart/middle-1}
 \addimg{0.360}{-0.91}{0.05}{flow-chart/middle-2}
 \addimg{0.420}{-0.91}{0.05}{flow-chart/middle-3}
 \addimg{0.480}{-0.91}{0.05}{flow-chart/middle-4}
 \addimg{0.300}{-0.97}{0.05}{flow-chart/middle-5}
 \addimg{0.360}{-0.97}{0.05}{flow-chart/middle-6}
 \addimg{0.420}{-0.97}{0.05}{flow-chart/middle-7}
 \addimg{0.480}{-0.97}{0.05}{flow-chart/middle-8}
 \addimg{0.300}{-1.03}{0.05}{flow-chart/middle-9}
 \addimg{0.360}{-1.03}{0.05}{flow-chart/middle-10}

 \addlabel{north west}{0}{0.275}{-1.03}{black}{\tiny Stamps of middle}
\fi

\ifdefined\flowchartstampcen
 \addimg{0.0180}{-0.91}{0.05}{flow-chart/center-1}
 \addimg{0.0780}{-0.91}{0.05}{flow-chart/center-2}
 \addimg{0.1380}{-0.91}{0.05}{flow-chart/center-3}
 \addimg{0.1980}{-0.91}{0.05}{flow-chart/center-4}
 \addimg{0.0180}{-0.97}{0.05}{flow-chart/center-5}
 \addimg{0.0780}{-0.97}{0.05}{flow-chart/center-6}
 \addimg{0.1380}{-0.97}{0.05}{flow-chart/center-7}
 \addimg{0.1980}{-0.97}{0.05}{flow-chart/center-8}
 \addimg{0.0180}{-1.03}{0.05}{flow-chart/center-9}
 \addimg{0.0780}{-1.03}{0.05}{flow-chart/center-10}
 \addimg{0.1380}{-1.03}{0.05}{flow-chart/center-11}
 \addimg{0.1980}{-1.03}{0.05}{flow-chart/center-12}

 \addlabel{north west}{0}{-0.01}{-1.03}{black}{\tiny Stamps of center}
\fi

\ifdefined\flowchartpsfcenter
   \addimg{0.15}{-1.385}{0.20}{flow-chart/center-psf-mrk}
 \else
   \addimg{0.15}{-1.385}{0.20}{flow-chart/center-psf}
\fi

\ifdefined\flowchartpsfmiddle
 \ifdefined\flowchartmarkstackcenter
   \addimg{0.41}{-1.4}{0.235}{flow-chart/middle-psf-mrk}
 \else
   \addimg{0.41}{-1.4}{0.235}{flow-chart/middle-psf}
 \fi
\fi

\ifdefined\flowchartpsfouter
 \ifdefined\flowchartmarkstackmiddle
   \addimg{0.69}{-1.42}{0.27}{flow-chart/outer-psf-mrk}
 \else
   \addimg{0.69}{-1.42}{0.27}{flow-chart/outer-psf}
 \fi
\fi

\ifdefined\flowchartmarkstackmiddle
 \ifdefined\flowchartpsfmiddle
   \addredline{green}{0.528}{-1.385}{0.685}{-1.274}
   \addredline{green}{0.528}{-1.154}{0.685}{-1.27}
 \fi
\fi
\ifdefined\flowchartmarkstackcenter
 \ifdefined\flowchartpsfcenter
   \addredline{blue}{0.25}{-1.172}{0.353}{-1.212}
   \addredline{blue}{0.25}{-1.37}{0.353}{-1.325}
 \fi
\fi

\ifdefined\flowchartpsfcenter
  \addlabel{north west}{0}{0.09}{-1.17}{black}{\tiny Center}
\fi
\ifdefined\flowchartpsfmiddle
  \addlabel{north west}{0}{0.35}{-1.155}{black}{\tiny Middle}
\fi
\ifdefined\flowchartpsfouter
 \addlabel{north west}{0}{0.64}{-1.14}{black}{\tiny Outer}
\fi

\ifdefined\flowchartfinalpsf
 \addimg{0.25}{-1.765}{0.29}{flow-chart/final}

 \addlabel{north west}{0}{0.16}{-1.46}{black}{\tiny United PSF}

\fi

 \ifdefined\flowchartfinalpsf
 \addimg{0.6}{-1.765}{0.29}{flow-chart/final-2d}

 \addlabel{north west}{0}{0.5}{-1.46}{black}{\tiny Projected PSF}

 \fi

  \begin{axis}[
    at={(-0.09\linewidth, -1.75\linewidth)},
    hide axis,
    colorbar,
    point meta max=28,
    point meta min=10,
    colorbar/width=0.03\linewidth,
    colormap/blackwhite,
    colorbar style={
      rotate=0,
      ylabel style={yshift=-0.2cm, rotate=180},
      ylabel={Surface Brightness (mag/arcsec$^2$)},
      yticklabel pos=left,
      yticklabel style={anchor=east},},
    ] {};
  \end{axis}

\end{tikzpicture}

%% file: tex/src/fig-radial-profile.tex
\begin{tikzpicture}

  \begin{axis}
    [
      xmin=10e-3,
      xmax=5,
      ymin=10e-2,
      ymax=10e6,
      xmode=log,
      ymode=log,
      xlabel={Distance form center [arcmin]},
      ylabel={Relative flux},
      width=\linewidth,
      height=\linewidth,
      legend pos=north east,
      enlarge x limits=false,
      enlarge y limits=false,
    ]

    \addplot [blue!80!white, line width=2pt]
    table [y expr=\thisrowno{1}*1200,
           x expr=\thisrowno{0}*0.555573/60]
    {tex/build/figures/radial-profile/center-rad.txt};
    \addlegendentry{Center part radial profile}

    \addplot [green!90!white, line width=3.5pt]
    table [y expr=\thisrowno{1}*70,
           x expr=\thisrowno{0}*0.555573/60]
    {tex/build/figures/radial-profile/middle-rad.txt};
    \addlegendentry{Middle part radial profile}

    \addplot [red!60!black,line width=4pt]
    table [x expr=\thisrowno{0}*0.555573/60]
    {tex/build/figures/radial-profile/outer-rad.txt};
    \addlegendentry{Outer part radial profile}

    \addplot[mark=none, blue!80!white, line width=1pt] coordinates {(0.09,0.00001) (0.09,10000000)};
    \addplot[mark=none, green!90!white, line width=1pt] coordinates {(0.25,0.00001) (0.25,10000000)};

  \end{axis}
\end{tikzpicture}

%% file: tex/src/fig-sat-obj-coord.tex
\begin{center}

\begin{tikzpicture}

       \node[anchor=north, line width=0.1mm,
         inner sep=0.2\pgflinewidth] (img) at (-0.8\linewidth,0.4\linewidth)
            {\includegraphics [width=0.498 \linewidth]
              {tex/build/figures/sat-obj-coord/bleed.pdf}};

   \begin{axis}
     [ at={(-1.047\linewidth,0.403\linewidth)},
       xmin=7,
       ymin=0.5,
       width=0.671\linewidth,
       height=0.41\linewidth,
       legend pos=north east,
       legend style={nodes={scale=0.4, transform shape}},
       enlarge x limits=false,
       xticklabel pos=upper,
       yticklabel pos=right,
       ticks=none,
     ]

     \addplot [fill=yellow,yellow!70!black,ultra thick] table {tex/build/figures/sat-obj-coord/x.txt};

     \end{axis}

     \begin{axis}
       [ at={(-0.435\linewidth,0.363\linewidth)},
         anchor=north,
         rotate around={270:(current axis.origin)},
         xmin=1550,
         xmax=5300,
         ymin=0.5,
         width=1.084\linewidth,
         height=0.47\linewidth,
         legend pos=north west,
         legend style={nodes={scale=0.4, transform shape}},
         enlarge x limits=false,
         yticklabel pos=left,
         xticklabel pos=top,
         ticks=none,
       ]

       \addplot [fill=yellow,yellow!70!black,ultra thick] table {tex/build/figures/sat-obj-coord/y.txt};

     \end{axis}

     \draw [gray,very thick, dotted] (-7.23,-4.6) -- (-7.23,5.7);
     \draw [gray,very thick, dotted] (-9.43,-0.51) -- (-2.29,-0.51);

\end{tikzpicture}
\end{center}

%% file: tex/src/fig-sort-bright-sat-star.tex
\begin{tikzpicture}

  \begin{axis}
    [
      ymax=5e4,
      xmode=log,
      ymode=log,
      xlabel={Distance form center [pix]},
      ylabel={Count},
      width=\linewidth,
      height=\linewidth,
      legend pos=north east,
      enlarge x limits=false,
      enlarge y limits=false,
    ]

    \addplot [gold!10!white,fill=black!10!white,fill opacity=0.5] coordinates
             {(18,0) (18,100000) (50,100000) (1000,0)}
             \closedcycle;

    \addplot [blueviolet!90!black, line width=2pt]
    table {tex/build/figures/sort-bright-sat-star/1-radial-profile-398.txt};

    \addplot [blue, line width=2pt]
    table {tex/build/figures/sort-bright-sat-star/2-radial-profile-619.txt};

    \addplot [green,line width=2pt]
    table {tex/build/figures/sort-bright-sat-star/3-radial-profile-448.txt};

    \addplot [orange,line width=2pt]
    table {tex/build/figures/sort-bright-sat-star/4-radial-profile-685.txt};

    \addplot [red,line width=2pt]
    table {tex/build/figures/sort-bright-sat-star/6-radial-profile-399.txt};

    \draw [black,very thick, dashed] (0,24000) -- (1000,24000);

  \end{axis}

  \node[anchor=east, draw=blueviolet, line width=0.5mm,
    inner sep=0.5\pgflinewidth] (img) at (0.145\linewidth,0.91\linewidth)
       {\includegraphics [width=0.135\linewidth]
         {tex/build/figures/sort-bright-sat-star/398-stamp.pdf}};

  \node[anchor=east, draw=blue, line width=0.5mm,
    inner sep=0.5\pgflinewidth] (img) at (0.315\linewidth,0.91\linewidth)
       {\includegraphics [width=0.135\linewidth]
         {tex/build/figures/sort-bright-sat-star/619-stamp.pdf}};

  \node[anchor=east, draw=green, line width=0.5mm,
    inner sep=0.5\pgflinewidth] (img) at (0.485\linewidth,0.91\linewidth)
       {\includegraphics [width=0.135\linewidth]
         {tex/build/figures/sort-bright-sat-star/448-stamp.pdf}};

  \node[anchor=east, draw=orange, line width=0.5mm,
    inner sep=0.5\pgflinewidth] (img) at (0.655\linewidth,0.91\linewidth)
       {\includegraphics [width=0.135\linewidth]
         {tex/build/figures/sort-bright-sat-star/685-stamp.pdf}};

  \node[anchor=east, draw=red, line width=0.5mm,
    inner sep=0.5\pgflinewidth] (img) at (0.825\linewidth,0.91\linewidth)
       {\includegraphics [width=0.135\linewidth]
         {tex/build/figures/sort-bright-sat-star/399-stamp.pdf}};

  \node[anchor=east,very thick] at (0.83\linewidth,0.78\linewidth)
       {Blank (saturation+nonlinearity) level};

\end{tikzpicture}

%% file: tex/src/fig-sat-ghost.tex
\begin{tikzpicture}

  \node [at={(0.13\linewidth,-0.025\linewidth)},
         rectangle,
         very thick,
         text centered,
         rounded corners,
         minimum width=0.9\linewidth,
         minimum height=0.55\linewidth,
         draw=white!50!black!50,
         fill=black!10!white!8!white] {};

  \draw[->, thick, rounded corners, line width=15pt] [orange!70!white]
       (-0.11\linewidth,0.09\linewidth)
       -- (-0.03\linewidth,0.09\linewidth);

  \draw[->, thick, rounded corners, line width=15pt] [orange!70!white]
       (0.195\linewidth,0.09\linewidth)
       -- (0.275\linewidth,0.09\linewidth);

  \draw[->, thick, rounded corners, line width=8pt] [orange!70!white]
       (0.46\linewidth,-0.002\linewidth)
       -- (0.46\linewidth,-0.0455\linewidth);

  \draw[->, thick, rounded corners, line width=15pt] [orange!70!white]
       (0\linewidth,-0.15\linewidth)
       -- (-0.075\linewidth,-0.15\linewidth);

  \draw[->, thick, rounded corners, line width=15pt] [orange!70!white]
       (0.3\linewidth,-0.15\linewidth)
       -- (0.22\linewidth,-0.15\linewidth);

  \begin{axis}
    [
      xmin=0,
      width=0.28\linewidth,
      height=0.28\linewidth,
      legend pos=north east,
      legend style={nodes={scale=0.7, transform shape}},
      enlarge x limits=false,
      xticklabel pos=upper,
      y dir=reverse,
      xlabel={Distance form center [arcsec]},
      ylabel={SB [mag/arcsec$^2$]},
    ]

    \addplot [red!70!black,ultra thick] table {tex/build/figures/sat-ghost/1097368-0-90-radial-profile.txt};
    \addlegendentry{Top right}
    \addplot [blue,ultra thick] table {tex/build/figures/sat-ghost/1097368-90-180-radial-profile.txt};
    \addlegendentry{Top left}
    \addplot [green!50!black,ultra thick] table {tex/build/figures/sat-ghost/1097368-180-270-radial-profile.txt};
    \addlegendentry{Bottem left}
    \addplot [violet!50!black,ultra thick] table {tex/build/figures/sat-ghost/1097368-270-360-radial-profile.txt};
    \addlegendentry{Bottem right}

  \end{axis}

  \node[anchor=north, line width=0.5mm,
        inner sep=0.2\pgflinewidth] (img) at (-0.2\linewidth,0.2\linewidth)
       {\includegraphics [width=0.2 \linewidth]
         {tex/build/figures/sat-ghost/1-raw.pdf}};

  \node[anchor=north, line width=0.5mm,
        inner sep=0.2\pgflinewidth] (img) at (0.4\linewidth,0.2\linewidth)
       {\includegraphics [width=0.2 \linewidth]
         {tex/build/figures/sat-ghost/1097368-prof2d-full.pdf}};

  \node[anchor=north, line width=0.5mm,
        inner sep=0.2\pgflinewidth] (img) at (-0.2\linewidth,-0.047\linewidth)
       {\includegraphics [width=0.2 \linewidth]
         {tex/build/figures/sat-ghost/1-stamp.pdf}};

  \node[anchor=north, line width=0.5mm,
        inner sep=0.2\pgflinewidth] (img) at (0.1\linewidth,-0.047\linewidth)
       {\includegraphics [width=0.2 \linewidth]
         {tex/build/figures/sat-ghost/1097368-nc-crop.pdf}};

  \node[anchor=north, line width=0.5mm,
        inner sep=0.2\pgflinewidth] (img) at (0.4\linewidth,-0.047\linewidth)
       {\includegraphics [width=0.2 \linewidth]
         {tex/build/figures/sat-ghost/1097368-subtraction.pdf}};

  \node[anchor=south, rotate=0] at (-0.138\linewidth,0.173\linewidth)
       {\textcolor{red!70!black}{\tiny\bf (Top right)}};

  \node[anchor=south, rotate=0] at (-0.268\linewidth,0.173\linewidth)
       {\textcolor{blue}{\tiny\bf (Top left)}};

  \node[anchor=south, rotate=0] at (-0.148\linewidth,-0.005\linewidth)
       {\textcolor{violet!50!black}{\tiny\bf (Bottom right)}};

  \node[anchor=south, rotate=0] at (-0.255\linewidth,-0.005\linewidth)
       {\textcolor{green!50!black}{\tiny\bf (Bottom left)}};

  \node[anchor=south, rotate=0] at (-0.2\linewidth,-0.04\linewidth)
       {\textcolor{orange!70!black}{\bf a}};

  \node[anchor=south, rotate=0] at (0.095\linewidth,-0.04\linewidth)
       {\textcolor{orange!70!black}{\bf b}};

  \node[anchor=south, rotate=0] at (0.4\linewidth,-0.04\linewidth)
       {\textcolor{orange!70!black}{\bf c}};

  \node[anchor=south, rotate=0] at (-0.2\linewidth,-0.285\linewidth)
       {\textcolor{orange!70!black}{\bf f}};

  \node[anchor=south, rotate=0] at (0.095\linewidth,-0.285\linewidth)
       {\textcolor{orange!70!black}{\bf e}};

  \node[anchor=south, rotate=0] at (0.4\linewidth,-0.285\linewidth)
       {\textcolor{orange!70!black}{\bf d}};

  \begin{axis}[
    at={(0.445\linewidth,-0.247\linewidth)},
    hide axis,
    colorbar,
    point meta max=28,
    point meta min=20,
    colorbar/width=0.01\linewidth,
    height=0.53\linewidth,
    colormap/blackwhite,
    colorbar style={
      rotate=0,
      ylabel style={yshift=-0.1cm, rotate=180},
      ylabel={Surface Brightness (mag/arcsec$^2$)},
      yticklabel pos=right,
      yticklabel style={anchor=west},},
    ] {};
  \end{axis}

\end{tikzpicture}

%% file: tex/src/fig-psf-subtract.tex
\begin{tikzpicture}

  \node[anchor=north, line width=0.5mm,
        inner sep=0.2\pgflinewidth] (img) at (-0.06\linewidth,0.15\linewidth)
       {\includegraphics [width=0.25\linewidth]
        {tex/build/figures/psf-sub/1066789.pdf}};

  \node[anchor=north, line width=0.5mm,
        inner sep=0.2\pgflinewidth] (img) at (0.22\linewidth,0.15\linewidth)
       {\includegraphics [width=0.25\linewidth]
         {tex/build/figures/psf-sub/1066789-scatterlight.pdf}};

  \node[anchor=north, line width=0.5mm,
        inner sep=0.2\pgflinewidth] (img) at (0.505\linewidth,0.15\linewidth)
       {\includegraphics [width=0.25\linewidth]
         {tex/build/figures/psf-sub/1066789-subtract.pdf}};

  \node[anchor=south, rotate=0] at (-0.06\linewidth,-0.1\linewidth)
       {\textcolor{black}{\bf\huge a}};

  \node[anchor=south, rotate=0] at (0.22\linewidth,-0.1\linewidth)
       {\textcolor{black}{\bf\huge b}};

  \node[anchor=south, scale=1] at (0.3\linewidth,-0.085\linewidth)
       {\textcolor{black}{\tiny 3 arcmin}};

  \node[anchor=south, rotate=0] at (0.505\linewidth,-0.1\linewidth)
       {\textcolor{black}{\bf\huge c}};

  \begin{axis}[
    at={(0.585\linewidth,-0.101\linewidth)},
    hide axis,
    colorbar,
    point meta max=30,
    point meta min=16,
    colorbar/width=0.01\linewidth,
    height=0.335\linewidth,
    colormap/blackwhite,
    colorbar style={
      rotate=0,
      ylabel style={yshift=-0.5cm, rotate=180, font=\tiny},
      ylabel={Surface Brightness (mag/arcsec$^2$)},
      yticklabel pos=right,
      yticklabel style={at={(0.585\linewidth,-0.101\linewidth)}, anchor=west, xshift=-0.05cm}},
    ] {};
  \end{axis}

  \begin{axis}[
    at={(0.3\linewidth,-0.101\linewidth)},
    hide axis,
    colorbar,
    point meta max=30,
    point meta min=16,
    colorbar/width=0.01\linewidth,
    height=0.335\linewidth,
    colormap/blackwhite,
    colorbar style={
      rotate=0,
      ylabel style={yshift=-0.1cm, rotate=180},
      yticklabel pos=right,
      yticklabel style={anchor=west},},
    ] {};
  \end{axis}

  \begin{axis}[
    at={(0.015\linewidth,-0.101\linewidth)},
    hide axis,
    colorbar,
    point meta max=30,
    point meta min=13,
    colorbar/width=0.01\linewidth,
    height=0.335\linewidth,
    colormap/blackwhite,
    colorbar style={
      rotate=0,
      ylabel style={yshift=-0.5cm, rotate=180},
      yticklabel pos=right,
      yticklabel style={anchor=west},},
    ] {};
  \end{axis}

\end{tikzpicture}

%% file: tex/src/fig-psf-slope.tex
\begin{tikzpicture}
  \begin{groupplot}[
      group style={group size=4 by 3,
                   horizontal sep=0pt,
                   vertical sep=0pt,
                   xticklabels at=edge bottom,
                   yticklabels at=edge left},
      xmin=0,
      ymin=0,
      ymax=25,
      xmax=5.8,
      xmode=log,
      no markers,
      y dir=reverse,
      height=0.23\linewidth,
      width=0.31\linewidth,
      legend pos=south west,
      ]

    \nextgroupplot
    \addplot [ujavacolor!70!black, line width=1pt]
    table [x expr=\thisrowno{0}*0.555573/60]
    {tex/build/figures/slope/1043079.txt};

    \addplot [ujavacolor!70!black, line width=1pt]
    table [x expr=\thisrowno{0}*0.555573/60]
    {tex/build/figures/slope/1043081.txt};

    \addplot [ujavacolor!70!black, line width=1pt]
    table [x expr=\thisrowno{0}*0.555573/60]
          {tex/build/figures/slope/1043083.txt};

    \addplot [ujavacolor, line width=1pt]
    table [x expr=\thisrowno{0}*0.555573/60]
          {tex/build/figures/slope/1053485.txt};

    \addplot [ujavacolor, line width=1pt]
    table [x expr=\thisrowno{0}*0.555573/60]
          {tex/build/figures/slope/1053486.txt};

    \addplot [ujavacolor, line width=1pt]
    table [x expr=\thisrowno{0}*0.555573/60]
          {tex/build/figures/slope/1053487.txt};

    \addplot [ujavacolor!40!white, line width=1pt]
    table [x expr=\thisrowno{0}*0.555573/60]
          {tex/build/figures/slope/1096979.txt};

    \addplot [ujavacolor!40!white, line width=1pt]
    table [x expr=\thisrowno{0}*0.555573/60]
          {tex/build/figures/slope/1096980.txt};

    \addplot [ujavacolor!40!white, line width=1pt]
    table [x expr=\thisrowno{0}*0.555573/60]
          {tex/build/figures/slope/1096983.txt};
    \addlegendentry{uJAVA}

    \nextgroupplot
    \addplot [j0378color!70!black, line width=1pt]
    table [x expr=\thisrowno{0}*0.555573/60]
          {tex/build/figures/slope/1043507.txt};

    \addplot [j0378color!70!black, line width=1pt]
    table [x expr=\thisrowno{0}*0.555573/60]
          {tex/build/figures/slope/1043510.txt};

    \addplot [j0378color!70!black, line width=1pt]
    table [x expr=\thisrowno{0}*0.555573/60]
          {tex/build/figures/slope/1043511.txt};

    \addplot [j0378color, line width=1pt]
    table [x expr=\thisrowno{0}*0.555573/60]
          {tex/build/figures/slope/1053932.txt};

    \addplot [j0378color, line width=1pt]
    table [x expr=\thisrowno{0}*0.555573/60]
          {tex/build/figures/slope/1053933.txt};

    \addplot [j0378color, line width=1pt]
    table [x expr=\thisrowno{0}*0.555573/60]
          {tex/build/figures/slope/1053934.txt};

    \addplot [j0378color!40!white, line width=1pt]
    table [x expr=\thisrowno{0}*0.555573/60]
          {tex/build/figures/slope/1097444.txt};

    \addplot [j0378color!40!white, line width=1pt]
    table [x expr=\thisrowno{0}*0.555573/60]
          {tex/build/figures/slope/1097445.txt};

    \addplot [j0378color!40!white, line width=1pt]
    table [x expr=\thisrowno{0}*0.555573/60]
          {tex/build/figures/slope/1097446.txt};
    \addlegendentry{J0378}

    \nextgroupplot
    \addplot [j0395color!70!black, line width=1pt]
    table [x expr=\thisrowno{0}*0.555573/60]
          {tex/build/figures/slope/1043731.txt};

    \addplot [j0395color!70!black, line width=1pt]
    table [x expr=\thisrowno{0}*0.555573/60]
          {tex/build/figures/slope/1043732.txt};

    \addplot [j0395color!70!black, line width=1pt]
    table [x expr=\thisrowno{0}*0.555573/60]
          {tex/build/figures/slope/1043733.txt};

    \addplot [j0395color, line width=1pt]
    table [x expr=\thisrowno{0}*0.555573/60]
          {tex/build/figures/slope/1054139.txt};

    \addplot [j0395color, line width=1pt]
    table [x expr=\thisrowno{0}*0.555573/60]
          {tex/build/figures/slope/1054140.txt};

    \addplot [j0395color, line width=1pt]
    table [x expr=\thisrowno{0}*0.555573/60]
          {tex/build/figures/slope/1054141.txt};

    \addplot [j0395color!40!white, line width=1pt]
    table [x expr=\thisrowno{0}*0.555573/60]
          {tex/build/figures/slope/1097683.txt};

    \addplot [j0395color!40!white, line width=1pt]
    table [x expr=\thisrowno{0}*0.555573/60]
          {tex/build/figures/slope/1097686.txt};

    \addplot [j0395color!40!white, line width=1pt]
    table [x expr=\thisrowno{0}*0.555573/60]
          {tex/build/figures/slope/1097690.txt};
    \addlegendentry{J0395}

    \nextgroupplot
    \addplot [j0410color!70!black, line width=1pt]
    table [x expr=\thisrowno{0}*0.555573/60]
          {tex/build/figures/slope/1043368.txt};

    \addplot [j0410color!70!black, line width=1pt]
    table [x expr=\thisrowno{0}*0.555573/60]
          {tex/build/figures/slope/1043369.txt};

    \addplot [j0410color!70!black, line width=1pt]
    table [x expr=\thisrowno{0}*0.555573/60]
          {tex/build/figures/slope/1043370.txt};

    \addplot [j0410color, line width=1pt]
    table [x expr=\thisrowno{0}*0.555573/60]
          {tex/build/figures/slope/1053778.txt};

    \addplot [j0410color, line width=1pt]
    table [x expr=\thisrowno{0}*0.555573/60]
          {tex/build/figures/slope/1053779.txt};

    \addplot [j0410color, line width=1pt]
    table [x expr=\thisrowno{0}*0.555573/60]
          {tex/build/figures/slope/1053781.txt};

    \addplot [j0410color!40!white, line width=1pt]
    table [x expr=\thisrowno{0}*0.555573/60]
          {tex/build/figures/slope/1097275.txt};

    \addplot [j0410color!40!white, line width=1pt]
    table [x expr=\thisrowno{0}*0.555573/60]
          {tex/build/figures/slope/1097277.txt};

    \addplot [j0410color!40!white, line width=1pt]
    table [x expr=\thisrowno{0}*0.555573/60]
          {tex/build/figures/slope/1097279.txt};
    \addlegendentry{J0410}

    \nextgroupplot
    \addplot [j0430color!70!black, line width=1pt]
    table [x expr=\thisrowno{0}*0.555573/60]
          {tex/build/figures/slope/1043152.txt};

    \addplot [j0430color!70!black, line width=1pt]
    table [x expr=\thisrowno{0}*0.555573/60]
          {tex/build/figures/slope/1043153.txt};

    \addplot [j0430color!70!black, line width=1pt]
    table [x expr=\thisrowno{0}*0.555573/60]
          {tex/build/figures/slope/1043155.txt};

    \addplot [j0430color, line width=1pt]
    table [x expr=\thisrowno{0}*0.555573/60]
          {tex/build/figures/slope/1053557.txt};

    \addplot [j0430color, line width=1pt]
    table [x expr=\thisrowno{0}*0.555573/60]
          {tex/build/figures/slope/1053558.txt};

    \addplot [j0430color, line width=1pt]
    table [x expr=\thisrowno{0}*0.555573/60]
          {tex/build/figures/slope/1053559.txt};

    \addplot [j0430color!40!white, line width=1pt]
    table [x expr=\thisrowno{0}*0.555573/60]
          {tex/build/figures/slope/1097050.txt};

    \addplot [j0430color!40!white, line width=1pt]
    table [x expr=\thisrowno{0}*0.555573/60]
          {tex/build/figures/slope/1097051.txt};

    \addplot [j0430color!40!white, line width=1pt]
    table [x expr=\thisrowno{0}*0.555573/60]
          {tex/build/figures/slope/1097052.txt};
    \addlegendentry{J0430}

    \nextgroupplot
    \addplot [gsdsscolor!70!black, line width=1pt]
    table [x expr=\thisrowno{0}*0.555573/60]
          {tex/build/figures/slope/1043651.txt};

    \addplot [gsdsscolor!70!black, line width=1pt]
    table [x expr=\thisrowno{0}*0.555573/60]
          {tex/build/figures/slope/1043652.txt};

    \addplot [gsdsscolor!70!black, line width=1pt]
    table [x expr=\thisrowno{0}*0.555573/60]
          {tex/build/figures/slope/1043653.txt};

    \addplot [gsdsscolor, line width=1pt]
    table [x expr=\thisrowno{0}*0.555573/60]
          {tex/build/figures/slope/1054064.txt};

    \addplot [gsdsscolor, line width=1pt]
    table [x expr=\thisrowno{0}*0.555573/60]
          {tex/build/figures/slope/1054067.txt};

    \addplot [gsdsscolor, line width=1pt]
    table [x expr=\thisrowno{0}*0.555573/60]
          {tex/build/figures/slope/1054070.txt};

    \addplot [gsdsscolor!40!white, line width=1pt]
    table [x expr=\thisrowno{0}*0.555573/60]
          {tex/build/figures/slope/1097603.txt};

    \addplot [gsdsscolor!40!white, line width=1pt]
    table [x expr=\thisrowno{0}*0.555573/60]
          {tex/build/figures/slope/1097604.txt};

    \addplot [gsdsscolor!40!white, line width=1pt]
    table [x expr=\thisrowno{0}*0.555573/60]
          {tex/build/figures/slope/1097605.txt};
    \addlegendentry{gSDSS}

    \nextgroupplot
    \addplot [j0515color!70!black, line width=1pt]
    table [x expr=\thisrowno{0}*0.555573/60]
          {tex/build/figures/slope/1043296.txt};

    \addplot [j0515color!70!black, line width=1pt]
    table [x expr=\thisrowno{0}*0.555573/60]
          {tex/build/figures/slope/1043297.txt};

    \addplot [j0515color!70!black, line width=1pt]
    table [x expr=\thisrowno{0}*0.555573/60]
          {tex/build/figures/slope/1043298.txt};

    \addplot [j0515color, line width=1pt]
    table [x expr=\thisrowno{0}*0.555573/60]
          {tex/build/figures/slope/1053707.txt};

    \addplot [j0515color, line width=1pt]
    table [x expr=\thisrowno{0}*0.555573/60]
          {tex/build/figures/slope/1053708.txt};

    \addplot [j0515color, line width=1pt]
    table [x expr=\thisrowno{0}*0.555573/60]
          {tex/build/figures/slope/1053709.txt};

    \addplot [j0515color!40!white, line width=1pt]
    table [x expr=\thisrowno{0}*0.555573/60]
          {tex/build/figures/slope/1097201.txt};

    \addplot [j0515color!40!white, line width=1pt]
    table [x expr=\thisrowno{0}*0.555573/60]
          {tex/build/figures/slope/1097203.txt};

    \addplot [j0515color!40!white, line width=1pt]
    table [x expr=\thisrowno{0}*0.555573/60]
          {tex/build/figures/slope/1097206.txt};
    \addlegendentry{J0515}

    \nextgroupplot
    \addplot [rsdsscolor!70!black, line width=1pt]
    table [x expr=\thisrowno{0}*0.555573/60]
          {tex/build/figures/slope/1043439.txt};

    \addplot [rsdsscolor!70!black, line width=1pt]
    table [x expr=\thisrowno{0}*0.555573/60]
          {tex/build/figures/slope/1043440.txt};

    \addplot [rsdsscolor!70!black, line width=1pt]
    table [x expr=\thisrowno{0}*0.555573/60]
          {tex/build/figures/slope/1043441.txt};

    \addplot [rsdsscolor, line width=1pt]
    table [x expr=\thisrowno{0}*0.555573/60]
          {tex/build/figures/slope/1053858.txt};

    \addplot [rsdsscolor, line width=1pt]
    table [x expr=\thisrowno{0}*0.555573/60]
          {tex/build/figures/slope/1053860.txt};

    \addplot [rsdsscolor, line width=1pt]
    table [x expr=\thisrowno{0}*0.555573/60]
          {tex/build/figures/slope/1053866.txt};

    %
    %
    \addlegendentry{rSDSS}

    \nextgroupplot
    \addplot [j0660color!70!black, line width=1pt]
    table [x expr=\thisrowno{0}*0.555573/60]
          {tex/build/figures/slope/1043574.txt};

    \addplot [j0660color!70!black, line width=1pt]
    table [x expr=\thisrowno{0}*0.555573/60]
          {tex/build/figures/slope/1043575.txt};

    \addplot [j0660color!70!black, line width=1pt]
    table [x expr=\thisrowno{0}*0.555573/60]
          {tex/build/figures/slope/1043576.txt};

    \addplot [j0660color, line width=1pt]
    table [x expr=\thisrowno{0}*0.555573/60]
          {tex/build/figures/slope/1053992.txt};

    \addplot [j0660color, line width=1pt]
    table [x expr=\thisrowno{0}*0.555573/60]
          {tex/build/figures/slope/1053993.txt};

    \addplot [j0660color, line width=1pt]
    table [x expr=\thisrowno{0}*0.555573/60]
          {tex/build/figures/slope/1053994.txt};

    \addplot [j0660color!40!white, line width=1pt]
    table [x expr=\thisrowno{0}*0.555573/60]
          {tex/build/figures/slope/1097519.txt};

    \addplot [j0660color!40!white, line width=1pt]
    table [x expr=\thisrowno{0}*0.555573/60]
          {tex/build/figures/slope/1097521.txt};

    \addplot [j0660color!40!white, line width=1pt]
    table [x expr=\thisrowno{0}*0.555573/60]
          {tex/build/figures/slope/1097524.txt};
    \addlegendentry{J0660}

    \nextgroupplot
    \addplot [isdsscolor!70!black, line width=1pt]
    table [x expr=\thisrowno{0}*0.555573/60]
          {tex/build/figures/slope/1043008.txt};

    \addplot [isdsscolor!70!black, line width=1pt]
    table [x expr=\thisrowno{0}*0.555573/60]
          {tex/build/figures/slope/1043010.txt};

    \addplot [isdsscolor!70!black, line width=1pt]
    table [x expr=\thisrowno{0}*0.555573/60]
          {tex/build/figures/slope/1043011.txt};

    \addplot [isdsscolor, line width=1pt]
    table [x expr=\thisrowno{0}*0.555573/60]
          {tex/build/figures/slope/1053419.txt};

    \addplot [isdsscolor, line width=1pt]
    table [x expr=\thisrowno{0}*0.555573/60]
          {tex/build/figures/slope/1053421.txt};


    \addplot [isdsscolor!40!white, line width=1pt]
    table [x expr=\thisrowno{0}*0.555573/60]
          {tex/build/figures/slope/1096900.txt};

    \addplot [isdsscolor!40!white, line width=1pt]
    table [x expr=\thisrowno{0}*0.555573/60]
          {tex/build/figures/slope/1096901.txt};

    \addplot [isdsscolor!40!white, line width=1pt]
    table [x expr=\thisrowno{0}*0.555573/60]
          {tex/build/figures/slope/1096902.txt};
    \addlegendentry{iSDSS}

    \nextgroupplot
    \addplot [j0861color!70!black, line width=1pt]
    table [x expr=\thisrowno{0}*0.555573/60]
          {tex/build/figures/slope/1043224.txt};

    \addplot [j0861color!70!black, line width=1pt]
    table [x expr=\thisrowno{0}*0.555573/60]
          {tex/build/figures/slope/1043225.txt};

    \addplot [j0861color!70!black, line width=1pt]
    table [x expr=\thisrowno{0}*0.555573/60]
          {tex/build/figures/slope/1043235.txt};

    \addplot [j0861color, line width=1pt]
    table [x expr=\thisrowno{0}*0.555573/60]
          {tex/build/figures/slope/1053633.txt};

    \addplot [j0861color, line width=1pt]
    table [x expr=\thisrowno{0}*0.555573/60]
          {tex/build/figures/slope/1053634.txt};

    \addplot [j0861color, line width=1pt]
    table [x expr=\thisrowno{0}*0.555573/60]
          {tex/build/figures/slope/1053641.txt};

    \addplot [j0861color!40!white, line width=1pt]
    table [x expr=\thisrowno{0}*0.555573/60]
          {tex/build/figures/slope/1097126.txt};

    \addplot [j0861color!40!white, line width=1pt]
    table [x expr=\thisrowno{0}*0.555573/60]
          {tex/build/figures/slope/1097127.txt};

    \addplot [j0861color!40!white, line width=1pt]
    table [x expr=\thisrowno{0}*0.555573/60]
          {tex/build/figures/slope/1097128.txt};
    \addlegendentry{J0861}

    \nextgroupplot
    \addplot [zsdsscolor!70!black, line width=1pt]
    table [x expr=\thisrowno{0}*0.555573/60]
          {tex/build/figures/slope/1043811.txt};

    \addplot [zsdsscolor!70!black, line width=1pt]
    table [x expr=\thisrowno{0}*0.555573/60]
          {tex/build/figures/slope/1043812.txt};

    \addplot [zsdsscolor!70!black, line width=1pt]
    table [x expr=\thisrowno{0}*0.555573/60]
          {tex/build/figures/slope/1043814.txt};

    \addplot [zsdsscolor, line width=1pt]
    table [x expr=\thisrowno{0}*0.555573/60]
          {tex/build/figures/slope/1054217.txt};

    \addplot [zsdsscolor, line width=1pt]
    table [x expr=\thisrowno{0}*0.555573/60]
          {tex/build/figures/slope/1054218.txt};

    \addplot [zsdsscolor, line width=1pt]
    table [x expr=\thisrowno{0}*0.555573/60]
          {tex/build/figures/slope/1054219.txt};

    \addplot [zsdsscolor!40!white, line width=1pt]
    table [x expr=\thisrowno{0}*0.555573/60]
          {tex/build/figures/slope/1097749.txt};

    \addplot [zsdsscolor!40!white, line width=1pt]
    table [x expr=\thisrowno{0}*0.555573/60]
          {tex/build/figures/slope/1097750.txt};

    \addplot [zsdsscolor!40!white, line width=1pt]
    table [x expr=\thisrowno{0}*0.555573/60]
          {tex/build/figures/slope/1097751.txt};
    \addlegendentry{zSDSS}

  \end{groupplot}

  \node[anchor=south] at (0.47\linewidth,-0.35\linewidth)
       {Distance from center [arcmin]};

  \node[anchor=south, rotate=90] at (-0.03\linewidth,-0.08\linewidth)
       {Surface brightness [mag/arcsec$^2$]};

\end{tikzpicture}

%% file: tex/src/fig-psf-depend-time.tex
\begin{tikzpicture}

  \begin{axis}
    [ at={(0\linewidth,0\linewidth)},
      ymin=16,
      ymax=27.7,
      xmin=0,
      xmax=8.1,
      xticklabel pos=upper,
      width=0.98\linewidth,
      height=0.78\linewidth,
      legend pos=north east,
      y dir=reverse,
      xtick=\empty,
      enlarge x limits=false,
      enlarge y limits=false,
    ]

    \addplot [gold!10!white,fill=black!10!white,fill opacity=0.5] coordinates
             {(0,16.04) (0.925955,16.04) (0.925955,27.65) (0,27.65)}
             \closedcycle;

    \addplot [gold!10!white,fill=black!10!white,fill opacity=0.5] coordinates
             {(1.85191,16.04) (2.777865,16.04) (2.777865,27.65) (1.85191,27.65)}
             \closedcycle;

    \addplot [gold!10!white,fill=black!10!white,fill opacity=0.5] coordinates
             {(3.70382,16.04) (4.629775,16.04) (4.629775,27.65) (3.70382,27.65)}
             \closedcycle;

    \addplot [gold!10!white,fill=black!10!white,fill opacity=0.5] coordinates
             {(5.55573,16.04) (6.481685,16.04) (6.481685,27.65) (5.55573,27.65)}
             \closedcycle;

    \addplot [gold!10!white,fill=black!10!white,fill opacity=0.5] coordinates
             {(7.40764,16.04) (8.333595,16.04) (8.333595,27.65) (7.40764,27.65)}
             \closedcycle;

    \addplot [only marks, red!90!black, mark size=1.5]
    table [y expr=\thisrowno{4},
           x expr=\thisrowno{0}*0.555573/60]
    {tex/build/figures/psf-depend-time/radial-profile-faint.txt};

    \addplot [only marks, blue!90!black, mark size=1.5]
    table [y expr=\thisrowno{4},
           x expr=\thisrowno{0}*0.555573/60]
    {tex/build/figures/psf-depend-time/radial-profile-bright.txt};

    \addplot[mark=none, gray!50!white, line width=0.5pt] coordinates {(2,14) (2,28)};
    \addplot[mark=none, gray!50!white, line width=0.5pt] coordinates {(4,14) (4,28)};
    \addplot[mark=none, gray!50!white, line width=0.5pt] coordinates {(6,14) (6,28)};

  \end{axis}

  \begin{axis}
    [ at={(0\linewidth,-0.179\linewidth)},
      ymin=-2,
      ymax=2,
      xmin=0,
      xmax=8.1,
      width=0.98\linewidth,
      height=0.35\linewidth,
      enlarge x limits=false,
      enlarge y limits=false,
    ]

    \addplot [only marks, black!30!white, mark size=0.65]
    table [y expr=\thisrowno{10},
           x expr=\thisrowno{0}*0.555573/60,
           x index=0, y index=10]
           {tex/build/figures/psf-depend-time/sb-diff-cat.txt};

    \addplot [only marks, black!80!black, mark size=1, error bars/.cd, y dir=both, y explicit]
    table [y expr=\thisrowno{1},
           x expr=\thisrowno{0}*0.555573/60,
           x index=0, y index=1, y error index=2] {tex/build/figures/psf-depend-time/cat.txt};

  \end{axis}

  \node[anchor=east, line width=0.5mm, draw=red,
    inner sep=0.5\pgflinewidth] (img) at (0.523\linewidth,0.47\linewidth)
       {\includegraphics [width=0.23\linewidth]
         {tex/build/figures/psf-depend-time/1043439-msk.pdf}};

  \node[anchor=east, line width=0.5mm, draw=blue,
    inner sep=0.5\pgflinewidth] (img) at (0.785\linewidth,0.47\linewidth)
       {\includegraphics [width=0.23\linewidth]
         {tex/build/figures/psf-depend-time/1066756-msk.pdf}};

  \node[anchor=south] at (0.4\linewidth,-0.28\linewidth)
       {Distance from center (arcmin)};

  \node[anchor=south, rotate=90] at (-0.08\linewidth,0.22\linewidth)
       {Surface Brightness (${mag/arcsec}^{2}$)};

\end{tikzpicture}

%% file: tex/src/fig-psf-depend-pos.tex
\begin{tikzpicture}

  \node[anchor=north, line width=0.5mm,
    inner sep=0.2\pgflinewidth] (img) at (-0.47\linewidth,0.2\linewidth)
       {\includegraphics [width=0.9\linewidth]
         {tex/build/figures/psf-depend-pos/1052126.pdf}};

  \node[anchor=south, scale=1] at (-0.13\linewidth,-0.7\linewidth)
       {\textcolor{black}{\bf \tiny 10 arcmin}};

  \node[anchor=north, draw=red, line width=0.5mm,
    inner sep=0.5\pgflinewidth] (img) at (-0.78\linewidth,-0.72\linewidth)
       {\includegraphics [width=0.27\linewidth]
         {tex/build/figures/psf-depend-pos/1-stamp.pdf}};

  \node[anchor=north, draw=yellow, line width=0.5mm,
    inner sep=0.5\pgflinewidth] (img) at (-0.47\linewidth,-0.72\linewidth)
       {\includegraphics [width=0.27\linewidth]
         {tex/build/figures/psf-depend-pos/2-stamp.pdf}};

  \node[anchor=north, draw=blue, line width=0.5mm,
    inner sep=0.5\pgflinewidth] (img) at (-0.16\linewidth,-0.72\linewidth)
       {\includegraphics [width=0.27\linewidth]
         {tex/build/figures/psf-depend-pos/3-stamp.pdf}};

  \begin{axis}
    [ at={(-0.909\linewidth,-1.245\linewidth)},
      ymin=-2,
      ymax=2,
      xmax=8,
      xlabel={Distance from center (arcmin)},
      width=1.06\linewidth,
      height=0.4\linewidth,
      legend pos=north east,
      enlarge x limits=false,
      enlarge y limits=false,
    ]

    \addplot [only marks, red!30!white, mark size=0.65]
    table [y expr=\thisrowno{10},
           x expr=\thisrowno{0}*0.555573/60,
           x index=0, y index=10]
           {tex/build/figures/psf-depend-pos/1-2-sub-sb-diff-cat.txt};

    \addplot [only marks, red!80!black, mark size=1, error bars/.cd, y dir=both, y explicit]
    table [y expr=\thisrowno{1},
           x expr=\thisrowno{0}*0.555573/60,
           x index=0, y index=1, y error index=2] {tex/build/figures/psf-depend-pos/1-2-sub.txt};

    \addplot [only marks, blue!30!white, mark size=0.65]
    table [y expr=\thisrowno{10},
           x expr=\thisrowno{0}*0.555573/60,
           x index=0, y index=10]
           {tex/build/figures/psf-depend-pos/3-2-sub-sb-diff-cat.txt};

    \addplot [only marks, blue!80!black, mark size=1, error bars/.cd, y dir=both, y explicit]
    table [y expr=\thisrowno{1},
           x expr=\thisrowno{0}*0.555573/60,
           x index=0, y index=1, y error index=2] {tex/build/figures/psf-depend-pos/3-2-sub.txt};

  \end{axis}
\end{tikzpicture}

%% file: tex/src/fig-psf-sky-estimate.tex
\begin{tikzpicture}

  \node[anchor=north, line width=0.5mm,
        inner sep=0.2\pgflinewidth] (img) at (-0.1\linewidth,0.2\linewidth)
       {\includegraphics [width=0.25\linewidth]
        {tex/build/figures/psf-sky-estimate/psfnotsub.pdf}};

  \node[anchor=north, line width=0.5mm,
        inner sep=0.2\pgflinewidth] (img) at (0.22\linewidth,0.2\linewidth)
       {\includegraphics [width=0.25\linewidth]
         {tex/build/figures/psf-sky-estimate/psfsub.pdf}};

  \node[anchor=north, line width=0.5mm,
        inner sep=0.2\pgflinewidth] (img) at (0.54\linewidth,0.2\linewidth)
       {\includegraphics [width=0.25\linewidth]
         {tex/build/figures/psf-sky-estimate/psfsub-sky-sub.pdf}};

  \node[anchor=south, rotate=0] at (-0.1\linewidth,-0.1\linewidth)
       {\textcolor{black}{\bf\large Sky map before PSF}};

  \node[anchor=south, rotate=0] at (0.22\linewidth,-0.1\linewidth)
       {\textcolor{black}{\bf\large Sky map after PSF}};

  \node[anchor=south, rotate=0] at (0.54\linewidth,-0.1\linewidth)
       {\textcolor{black}{\bf\large Subtraction}};

  \node[anchor=south, scale=0.7] at (0.635\linewidth,-0.042\linewidth)
       {\textcolor{green}{\tiny 3 arcmin}};

  \begin{axis}[
    at={(0.63\linewidth,-0.051\linewidth)},
    hide axis,
    colorbar,
    point meta max=25.5,
    point meta min=20.4,
    colorbar/width=0.01\linewidth,
    height=0.335\linewidth,
    colormap/blackwhite,
    colorbar style={
      rotate=0,
      ylabel style={yshift=-0.5cm, rotate=180},
      ylabel={Surface Brightness (mag/arcsec$^2$)},
      yticklabel pos=right,
      yticklabel style={anchor=west},},
    ] {};
  \end{axis}

  \begin{axis}[
    at={(0.31\linewidth,-0.051\linewidth)},
    hide axis,
    colorbar,
    point meta max=14.95,
    point meta min=14.9,
    colorbar/width=0.01\linewidth,
    height=0.335\linewidth,
    colormap/viridis,
    colorbar style={
      rotate=0,
      ylabel style={yshift=-0.1cm, rotate=180},
      yticklabel pos=right,
      yticklabel style={anchor=west},},
    ] {};
  \end{axis}

  \begin{axis}[
    at={(-0.01\linewidth,-0.051\linewidth)},
    hide axis,
    colorbar,
    point meta max=14.95,
    point meta min=14.9,
    colorbar/width=0.01\linewidth,
    height=0.335\linewidth,
    colormap/viridis,
    colorbar style={
      rotate=0,
      ylabel style={yshift=-0.1cm, rotate=180},
      yticklabel pos=right,
      yticklabel style={anchor=west},},
    ] {};
  \end{axis}

\end{tikzpicture}

%% file: tex/src/fig-ngc-4212.tex
\begin{tikzpicture}[spy using outlines=
    {rectangle, magnification=2, connect spies}]

    \begin{axis}
    [ at={(0\linewidth,0\linewidth)},
      ymin=16,
      ymax=23,
      xmax=7,
      xlabel={Distance from center (arcmin)},
      ylabel={Surface Brightness (mag/arcsec$^2$)},
      legend style={nodes={scale=1, transform shape}},
      y dir=reverse,
      width=0.98\linewidth,
      height=0.65\linewidth,
      legend pos=north east,
      enlarge x limits=false,
      enlarge y limits=false,
    ]

    \addplot [only marks, red!75!purple, mark size=0.85]
    table [y expr=\thisrowno{3},
           x expr=\thisrowno{0}*0.555573/60,
           x index=0, y index=3]
    {tex/build/figures/ngc-4212/crop-sb.txt};
    \addlegendentry{Bright stars turn on}

    \addplot [only marks, teal!85!green, mark size=0.8]
    table [y expr=\thisrowno{3},
           x expr=\thisrowno{0}*0.555573/60,
           x index=0, y index=3]
    {tex/build/figures/ngc-4212/psf-sub-sb.txt};
    \addlegendentry{Bright stars turn off}

     \coordinate (spypoint) at (1.26,17.4);
     \coordinate (magnifyglass) at (1.68,21.2);

    \spy [black, width=3.3cm, height=1.8cm] on (spypoint)
    in node[fill=white] at (magnifyglass);

    \end{axis}

\end{tikzpicture}

%% file: tex/src/fig-spike-angle.tex
\begin{center}
  \begin{tikzpicture}

  \node[anchor=east] (img) at (-0.016\linewidth,-0.12\linewidth)
       {\includegraphics [width=0.8\linewidth,height=0.46\linewidth]
         {tex/build/figures/spike-angle/polar.pdf}};

  \begin{axis}
    [ at={(-0.833\linewidth,-0.3\linewidth)},
      xticklabel pos=right,
      grid=minor,
      xlabel={Azimuthal angle},
      ylabel={Radius},
      ymin=250,
      yticklabels={},
      xmax=360,
      width=0.974\linewidth,
      height=0.57\linewidth,
      legend pos=north east,
      enlarge x limits=false,
      enlarge y limits=false,
    ]

    \addplot [red, thick]
    table {tex/build/figures/spike-angle/collapse.txt};

  \end{axis}

  \node[anchor=east] (img) at (-0.02\linewidth,-0.706\linewidth)
       {\includegraphics [width=0.8\linewidth]
         {tex/build/figures/spike-angle/1-stamp.pdf}};

\end{tikzpicture}
\end{center}

%% file: tex/build/macros/dependencies.tex
 
This research was done with the following free software programs and libraries: Bzip2 1.0.8, CFITSIO 4.5.0, CMake 3.31.5, cURL 8.11.1, Dash 0.5.12, Discoteq flock 0.4.0, Expat 2.6.4, File 5.46, Fontconfig 2.16.0, FreeType 2.13.3, Git 2.48.1, GNU Astronomy Utilities 0.22.90-ccb84 \citep{gnuastro}, GNU Autoconf 2.72, GNU Automake 1.17, GNU AWK 5.3.1, GNU Bash 5.2.37, GNU Binutils 2.43.1, GNU Bison 3.8.2, GNU Compiler Collection (GCC) 14.2.0, GNU Coreutils 9.6, GNU Diffutils 3.10, GNU Emacs 28.1, GNU Findutils 4.10.0, GNU gettext 0.23.1, GNU gperf 3.1, GNU Grep 3.11, GNU Gzip 1.13, GNU Integer Set Library 0.27, GNU libiconv 1.18, GNU Libtool 2.5.4, GNU libunistring 1.3, GNU M4 1.4.19, GNU Make 4.4.1, GNU Multiple Precision Arithmetic Library 6.3.0, GNU Multiple Precision Complex library, GNU Multiple Precision Floating-Point Reliably 4.2.1, GNU Nano 8.3, GNU NCURSES 6.5, GNU Readline 8.2.13, GNU Scientific Library 2.8, GNU Sed 4.9, GNU Tar 1.35, GNU Texinfo 7.2, GNU Wget 1.25.0, GNU Which 2.23, GPL Ghostscript 10.04.0, Help2man , Less 668, Libffi 3.4.7, Libgit2 1.9.0, libICE 1.1.2, Libidn 1.42, Libjpeg 9f, Libpaper 1.1.29, Libpng 1.6.46, libpthread-stubs (Xorg) 0.5, libSM 1.2.5, Libtiff 4.7.0, libXau (Xorg) 1.0.12, libxcb (Xorg) 1.17.0, libXdmcp (Xorg) 1.1.5, libXext 1.3.6, Libxml2 2.13.5, libXt 1.3.1, Lzip 1.25, OpenSSL 3.4.0, PatchELF 0.13, Perl 5.40.1, pkg-config 0.29.2, podlators 6.0.2, Python 3.13.2, util-Linux 2.40.4, util-macros (Xorg) 1.20.2, WCSLIB 8.4, X11 library 1.8, XCB-proto (Xorg) 1.17.0, xorgproto 2024.1, xtrans (Xorg) 1.5.2, XZ Utils 5.6.3 and Zlib 1.3.1. 
The \LaTeX{} source of the paper was compiled to make the PDF using the following packages: biber 2.20, biblatex 3.20, caption 68425 (revision), courier 61719 (revision), csquotes 5.2o, datetime 2.60, fancyvrb 4.5c, fmtcount 3.10, fontaxes 2.0.1, footmisc 7.0b, fp 2.1d, helvetic 61719 (revision), kastrup 15878 (revision), logreq 1.0, mweights 53520 (revision), newtx 1.756, pgf 3.1.10, pgfplots 1.18.1, preprint 2011, setspace 6.7b, sttools 3.1, texgyre 2.501, times 61719 (revision), titlesec 2.17, trimspaces 1.1, txfonts 15878 (revision), ulem 53365 (revision), xcolor 3.02, xkeyval 2.9, xpatch 0.3 and xstring 1.86. 
We are very grateful to all their creators for freely  providing this necessary infrastructure. This research  (and many other projects) would not be possible without  them.

%% file: paper.bbl
\begin{thebibliography}{54}
\expandafter\ifx\csname natexlab\endcsname\relax\def\natexlab#1{#1}\fi

\bibitem[{{Akhlaghi}(2019{\natexlab{a}})}]{akhlaghi19}
{Akhlaghi}, M. 2019{\natexlab{a}}, IAU Symposium 335, arXiv:1909.11230

\bibitem[{{Akhlaghi}(2019{\natexlab{b}})}]{akhlaghi19b}
{Akhlaghi}, M. 2019{\natexlab{b}}, in Astronomical Society of the Pacific
  Conference Series, Vol. 521, Astronomical Data Analysis Software and Systems
  XXVI, ed. M.~{Molinaro}, K.~{Shortridge}, \& F.~{Pasian}, 299

\bibitem[{{Akhlaghi}(2024)}]{gnuastrobook}
{Akhlaghi}, M. 2024, GNU Astronomy Utilities (version 0.23),
  https://doi.org/10.5281/zenodo.12738457 (Free Software Foundation)

\bibitem[{{Akhlaghi} \& {Ichikawa}(2015)}]{gnuastro}
{Akhlaghi}, M. \& {Ichikawa}, T. 2015, ApJS, 220, 1

\bibitem[{{Akhlaghi} {et~al.}(2021){Akhlaghi}, {Infante-Sainz}, {Roukema},
  {Khellat}, {Valls-Gabaud}, \& {Baena-Galle}}]{maneage}
{Akhlaghi}, M., {Infante-Sainz}, R., {Roukema}, B.~F., {et~al.} 2021, CiSE, 23,
  82

\bibitem[{{Anderson} \& {King}(2000)}]{anderson00}
{Anderson}, J. \& {King}, I.~R. 2000, \pasp, 112, 1360

\bibitem[{{Bazkiaei} {et~al.}(2024){Bazkiaei}, {Kelvin}, {Brough}, {O'Toole},
  {Watkins}, \& {Schmitz}}]{amir24}
{Bazkiaei}, A.~E., {Kelvin}, L.~S., {Brough}, S., {et~al.} 2024, in Society of
  Photo-Optical Instrumentation Engineers (SPIE) Conference Series, Vol. 13101,
  Software and Cyberinfrastructure for Astronomy VIII, ed. J.~{Ibsen} \&
  G.~{Chiozzi}, 131013N

\bibitem[{{Benitez} {et~al.}(2014){Benitez}, {Dupke}, {Moles}, {Sodre},
  {Cenarro}, {Marin-Franch}, {Taylor}, {Cristobal}, {Fernandez-Soto}, {Mendes
  de Oliveira}, {Cepa-Nogue}, {Abramo}, {Alcaniz}, {Overzier},
  {Hernandez-Monteagudo}, {Alfaro}, {Kanaan}, {Carvano}, {Reis}, {Martinez
  Gonzalez}, {Ascaso}, {Ballesteros}, {Xavier}, {Varela}, {Ederoclite},
  {Vazquez Ramio}, {Broadhurst}, {Cypriano}, {Angulo}, {Diego}, {Zandivarez},
  {Diaz}, {Melchior}, {Umetsu}, {Spinelli}, {Zitrin}, {Coe}, {Yepes}, {Vielva},
  {Sahni}, {Marcos-Caballero}, {Kitaura}, {Maroto}, {Masip}, {Tsujikawa},
  {Carneiro}, {Gonzalez Nuevo}, {Carvalho}, {Reboucas}, {Carvalho}, {Abdalla},
  {Bernui}, {Pigozzo}, {Ferreira}, {Chandrachani Devi}, {Bengaly}, {Campista},
  {Amorim}, {Asari}, {Bongiovanni}, {Bonoli}, {Bruzual}, {Cardiel}, {Cava},
  {Cid Fernandes}, {Coelho}, {Cortesi}, {Delgado}, {Diaz Garcia}, {Espinosa},
  {Galliano}, {Gonzalez-Serrano}, {Falcon-Barroso}, {Fritz}, {Fernandes},
  {Gorgas}, {Hoyos}, {Jimenez-Teja}, {Lopez-Aguerri}, {Lopez-San Juan},
  {Mateus}, {Molino}, {Novais}, {OMill}, {Oteo}, {Perez-Gonzalez}, {Poggianti},
  {Proctor}, {Ricciardelli}, {Sanchez-Blazquez}, {Storchi-Bergmann}, {Telles},
  {Schoennell}, {Trujillo}, {Vazdekis}, {Viironen}, {Daflon},
  {Aparicio-Villegas}, {Rocha}, {Ribeiro}, {Borges}, {Martins}, {Marcolino},
  {Martinez-Delgado}, {Perez-Torres}, {Siffert}, {Calvao}, {Sako}, {Kessler},
  {Alvarez-Candal}, {De Pra}, {Roig}, {Lazzaro}, {Gorosabel}, {Lopes de
  Oliveira}, {Lima-Neto}, {Irwin}, {Liu}, {Alvarez}, {Balmes}, {Chueca},
  {Costa-Duarte}, {da Costa}, {Dantas}, {Diaz}, {Fabregat}, {Ferrari},
  {Gavela}, {Gracia}, {Gruel}, {Gutierrez}, {Guzman}, {Hernandez-Fernandez},
  {Herranz}, {Hurtado-Gil}, {Jablonsky}, {Laporte}, {Le Tiran}, {Licandro},
  {Lima}, {Martin}, {Martinez}, {Montero}, {Penteado}, {Pereira}, {Peris},
  {Quilis}, {Sanchez-Portal}, {Soja}, {Solano}, {Torra}, \&
  {Valdivielso}}]{benitez14}
{Benitez}, N., {Dupke}, R., {Moles}, M., {et~al.} 2014, arXiv e-prints,
  arXiv:1403.5237

\bibitem[{{Bertin} \& {Arnouts}(1996)}]{sextractor}
{Bertin}, E. \& {Arnouts}, S. 1996, \aaps, 117, 393

\bibitem[{{Bonoli} {et~al.}(2021){Bonoli}, {Mar{\'\i}n-Franch}, {Varela},
  {V{\'a}zquez Rami{\'o}}, {Abramo}, {Cenarro}, {Dupke}, {V{\'\i}lchez},
  {Crist{\'o}bal-Hornillos}, {Gonz{\'a}lez Delgado},
  {Hern{\'a}ndez-Monteagudo}, {L{\'o}pez-Sanjuan}, {Muniesa}, {Civera},
  {Ederoclite}, {Hern{\'a}n-Caballero}, {Marra}, {Baqui}, {Cortesi},
  {Cypriano}, {Daflon}, {de Amorim}, {D{\'\i}az-Garc{\'\i}a}, {Diego},
  {Mart{\'\i}nez-Solaeche}, {P{\'e}rez}, {Placco}, {Prada}, {Queiroz},
  {Alcaniz}, {Alvarez-Candal}, {Cepa}, {Maroto}, {Roig}, {Siffert}, {Taylor},
  {Benitez}, {Moles}, {Sodr{\'e}}, {Carneiro}, {Mendes de Oliveira}, {Abdalla},
  {Angulo}, {Aparicio Resco}, {Balaguera-Antol{\'\i}nez}, {Ballesteros},
  {Brito-Silva}, {Broadhurst}, {Carrasco}, {Castro}, {Cid Fernandes}, {Coelho},
  {de Melo}, {Doubrawa}, {Fernandez-Soto}, {Ferrari}, {Finoguenov},
  {Garc{\'\i}a-Benito}, {Iglesias-P{\'a}ramo}, {Jim{\'e}nez-Teja}, {Kitaura},
  {Laur}, {Lopes}, {Lucatelli}, {Mart{\'\i}nez}, {Maturi}, {Overzier},
  {Pigozzo}, {Quartin}, {Rodr{\'\i}guez-Mart{\'\i}n}, {Salzano}, {Tamm},
  {Tempel}, {Umetsu}, {Valdivielso}, {von Marttens}, {Zitrin},
  {D{\'\i}az-Mart{\'\i}n}, {L{\'o}pez-Alegre}, {L{\'o}pez-Sainz},
  {Yanes-D{\'\i}az}, {Rueda-Teruel}, {Rueda-Teruel}, {Abril Iba{\~n}ez}, {L
  Ant{\'o}n Bravo}, {Bello Ferrer}, {Bielsa}, {Casino}, {Castillo}, {Chueca},
  {Cuesta}, {Garzar{\'a}n Calderaro}, {Iglesias-Marzoa}, {{\'I}niguez},
  {Lamadrid Gutierrez}, {Lopez-Martinez}, {Lozano-P{\'e}rez}, {Ma{\'\i}cas
  Sacrist{\'a}n}, {Molina-Ib{\'a}{\~n}ez}, {Moreno-Signes}, {Rodr{\'\i}guez
  Llano}, {Royo Navarro}, {Tilve Rua}, {Andrade}, {Alfaro}, {Akras},
  {Arnalte-Mur}, {Ascaso}, {Barbosa}, {Beltr{\'a}n Jim{\'e}nez}, {Benetti},
  {Bengaly}, {Bernui}, {Blanco-Pillado}, {Borges Fernandes}, {Bregman},
  {Bruzual}, {Calderone}, {Carvano}, {Casarini}, {Chaves-Montero},
  {Chies-Santos}, {Coutinho de Carvalho}, {Dimauro}, {Duarte Puertas},
  {Figueruelo}, {Gonz{\'a}lez-Serrano}, {Guerrero}, {Gurung-L{\'o}pez},
  {Herranz}, {Huertas-Company}, {Irwin}, {Izquierdo-Villalba}, {Kanaan},
  {Kehrig}, {Kirkpatrick}, {Lim}, {Lopes}, {Lopes de Oliveira},
  {Marcos-Caballero}, {Mart{\'\i}nez-Delgado}, {Mart{\'\i}nez-Gonz{\'a}lez},
  {Mart{\'\i}nez-Somonte}, {Oliveira}, {Orsi}, {Penna-Lima}, {Reis}, {Spinoso},
  {Tsujikawa}, {Vielva}, {Vitorelli}, {Xia}, {Yuan}, {Arroyo-Polonio},
  {Dantas}, {Galarza}, {Gon{\c{c}}alves}, {Gon{\c{c}}alves}, {Gonzalez},
  {Gonzalez}, {Greisel}, {Jim{\'e}nez-Esteban}, {Landim}, {Lazzaro}, {Magris},
  {Monteiro-Oliveira}, {Pereira}, {Rebou{\c{c}}as}, {Rodriguez-Espinosa},
  {Santos da Costa}, \& {Telles}}]{minijpas}
{Bonoli}, S., {Mar{\'\i}n-Franch}, A., {Varela}, J., {et~al.} 2021, \aap, 653,
  A31

\bibitem[{{Borlaff} {et~al.}(2022){Borlaff}, {G{\'o}mez-Alvarez}, {Altieri},
  {Marcum}, {Vavrek}, {Laureijs}, {Kohley}, {Buitrago}, {Cuillandre}, {Duc},
  {Gaspar Venancio}, {Amara}, {Andreon}, {Auricchio}, {Azzollini},
  {Baccigalupi}, {Balaguera-Antol{\'\i}nez}, {Baldi}, {Bardelli}, {Bender},
  {Biviano}, {Bodendorf}, {Bonino}, {Bozzo}, {Branchini}, {Brescia},
  {Brinchmann}, {Burigana}, {Cabanac}, {Camera}, {Candini}, {Capobianco},
  {Cappi}, {Carbone}, {Carretero}, {Carvalho}, {Casas}, {Castander},
  {Castellano}, {Castignani}, {Cavuoti}, {Cimatti}, {Cledassou},
  {Colodro-Conde}, {Congedo}, {Conselice}, {Conversi}, {Copin}, {Corcione},
  {Coupon}, {Courtois}, {Cropper}, {Da Silva}, {Degaudenzi}, {Di Ferdinando},
  {Douspis}, {Dubath}, {Duncan}, {Dupac}, {Dusini}, {Ealet}, {Fabricius},
  {Farina}, {Farrens}, {Ferreira}, {Ferriol}, {Finelli}, {Flose-Reimberg},
  {Fosalba}, {Frailis}, {Franceschi}, {Fumana}, {Galeotta}, {Ganga}, {Garilli},
  {Gillis}, {Giocoli}, {Gozaliasl}, {Graci{\'a}-Carpio}, {Grazian}, {Grupp},
  {Haugan}, {Holmes}, {Hormuth}, {Jahnke}, {Keihanen}, {Kermiche}, {Kiessling},
  {Kilbinger}, {Kirkpatrick}, {Kitching}, {Knapen}, {Kubik}, {K{\"u}mmel},
  {Kunz}, {Kurki-Suonio}, {Liebing}, {Ligori}, {Lilje}, {Lindholm}, {Lloro},
  {Mainetti}, {Maino}, {Mansutti}, {Marggraf}, {Markovic}, {Martinelli},
  {Martinet}, {Mart{\'\i}nez-Delgado}, {Marulli}, {Massey}, {Maturi},
  {Maurogordato}, {Medinaceli}, {Mei}, {Meneghetti}, {Merlin}, {Metcalf},
  {Meylan}, {Moresco}, {Morgante}, {Moscardini}, {Munari}, {Nakajima},
  {Neissner}, {Niemi}, {Nightingale}, {Nucita}, {Padilla}, {Paltani}, {Pasian},
  {Patrizii}, {Pedersen}, {Percival}, {Pettorino}, {Pires}, {Poncet}, {Popa},
  {Potter}, {Pozzetti}, {Raison}, {Rebolo}, {Renzi}, {Rhodes}, {Riccio},
  {Romelli}, {Roncarelli}, {Rosset}, {Rossetti}, {Saglia}, {S{\'a}nchez},
  {Sapone}, {Sauvage}, {Schneider}, {Scottez}, {Secroun}, {Seidel}, {Serrano},
  {Sirignano}, {Sirri}, {Skottfelt}, {Stanco}, {Starck}, {Sureau},
  {Tallada-Cresp{\'\i}}, {Taylor}, {Tenti}, {Tereno}, {Teyssier},
  {Toledo-Moreo}, {Torradeflot}, {Tutusaus}, {Valentijn}, {Valenziano},
  {Valiviita}, {Vassallo}, {Viel}, {Wang}, {Weller}, {Whittaker}, {Zacchei},
  {Zamorani}, \& {Zucca}}]{borlaff22}
{Borlaff}, A.~S., {G{\'o}mez-Alvarez}, P., {Altieri}, B., {et~al.} 2022, \aap,
  657, A92

\bibitem[{{Cenarro} {et~al.}(2019){Cenarro}, {Moles},
  {Crist{\'o}bal-Hornillos}, {Mar{\'\i}n-Franch}, {Ederoclite}, {Varela},
  {L{\'o}pez-Sanjuan}, {Hern{\'a}ndez-Monteagudo}, {Angulo}, {V{\'a}zquez
  Rami{\'o}}, {Viironen}, {Bonoli}, {Orsi}, {Hurier}, {San Roman}, {Greisel},
  {Vilella-Rojo}, {D{\'\i}az-Garc{\'\i}a}, {Logro{\~n}o-Garc{\'\i}a},
  {Gurung-L{\'o}pez}, {Spinoso}, {Izquierdo-Villalba}, {Aguerri}, {Allende
  Prieto}, {Bonatto}, {Carvano}, {Chies-Santos}, {Daflon}, {Dupke},
  {Falc{\'o}n-Barroso}, {Gon{\c{c}}alves}, {Jim{\'e}nez-Teja}, {Molino},
  {Placco}, {Solano}, {Whitten}, {Abril}, {Ant{\'o}n}, {Bello}, {Bielsa de
  Toledo}, {Castillo-Ram{\'\i}rez}, {Chueca}, {Civera},
  {D{\'\i}az-Mart{\'\i}n}, {Dom{\'\i}nguez-Mart{\'\i}nez},
  {Garzar{\'a}n-Calderaro}, {Hern{\'a}ndez-Fuertes}, {Iglesias-Marzoa},
  {I{\~n}iguez}, {Jim{\'e}nez Ruiz}, {Kruuse}, {Lamadrid}, {Lasso-Cabrera},
  {L{\'o}pez-Alegre}, {L{\'o}pez-Sainz}, {Ma{\'\i}cas}, {Moreno-Signes},
  {Muniesa}, {Rodr{\'\i}guez-Llano}, {Rueda-Teruel}, {Rueda-Teruel},
  {Soriano-Lagu{\'\i}a}, {Tilve}, {Valdivielso}, {Yanes-D{\'\i}az}, {Alcaniz},
  {Mendes de Oliveira}, {Sodr{\'e}}, {Coelho}, {Lopes de Oliveira}, {Tamm},
  {Xavier}, {Abramo}, {Akras}, {Alfaro}, {Alvarez-Candal}, {Ascaso}, {Beasley},
  {Beers}, {Borges Fernandes}, {Bruzual}, {Buzzo}, {Carrasco}, {Cepa},
  {Cortesi}, {Costa-Duarte}, {De Pr{\'a}}, {Favole}, {Galarza}, {Galbany},
  {Garcia}, {Gonz{\'a}lez Delgado}, {Gonz{\'a}lez-Serrano},
  {Guti{\'e}rrez-Soto}, {Hernandez-Jimenez}, {Kanaan}, {Kuncarayakti},
  {Landim}, {Laur}, {Licandro}, {Lima Neto}, {Lyman}, {Ma{\'\i}z
  Apell{\'a}niz}, {Miralda-Escud{\'e}}, {Morate}, {Nogueira-Cavalcante},
  {Novais}, {Oncins}, {Oteo}, {Overzier}, {Pereira}, {Rebassa-Mansergas},
  {Reis}, {Roig}, {Sako}, {Salvador-Rusi{\~n}ol}, {Sampedro},
  {S{\'a}nchez-Bl{\'a}zquez}, {Santos}, {Schmidtobreick}, {Siffert}, {Telles},
  \& {Vilchez}}]{cenarro19}
{Cenarro}, A.~J., {Moles}, M., {Crist{\'o}bal-Hornillos}, D., {et~al.} 2019,
  \aap, 622, A176

\bibitem[{{Comer{\'o}n} {et~al.}(2018){Comer{\'o}n}, {Salo}, \&
  {Knapen}}]{comeron18}
{Comer{\'o}n}, S., {Salo}, H., \& {Knapen}, J.~H. 2018, \aap, 610, A5

\bibitem[{{Coupon} {et~al.}(2018){Coupon}, {Czakon}, {Bosch}, {Komiyama},
  {Medezinski}, {Miyazaki}, \& {Oguri}}]{coupon18}
{Coupon}, J., {Czakon}, N., {Bosch}, J., {et~al.} 2018, \pasj, 70, S7

\bibitem[{{Crist{\'o}bal-Hornillos} {et~al.}(2012){Crist{\'o}bal-Hornillos},
  {Gruel}, {Varela}, {L{\'o}pez-Sainz}, {Ederoclite}, {Moles}, {Cenarro},
  {Mar{\'{\i}}n-Franch}, {Hern{\'a}ndez-Fuertes}, {Yanes-D{\'{\i}}az},
  {Chueca}, {Rueda-Teruel}, {Rueda-Teruel}, \& {Luis-Simoes}}]{upad}
{Crist{\'o}bal-Hornillos}, D., {Gruel}, N., {Varela}, J., {et~al.} 2012, in
  SPIE CS, Vol. 8451

\bibitem[{{de Jong}(2008)}]{dejong08}
{de Jong}, R.~S. 2008, \mnras, 388, 1521

\bibitem[{{de Vaucouleurs}(1958)}]{devacouleurs58}
{de Vaucouleurs}, G. 1958, \apj, 128, 465

\bibitem[{{Eskandarlou} \& {Akhlaghi}(2024)}]{Eskandarlou24}
{Eskandarlou}, S. \& {Akhlaghi}, M. 2024, Research Notes of the American
  Astronomical Society, 8, 168

\bibitem[{{Gaia Collaboration} {et~al.}(2023){Gaia Collaboration}, {Vallenari},
  {Brown}, {Prusti}, {de Bruijne}, {Arenou}, {Babusiaux}, {Biermann},
  {Creevey}, {Ducourant}, {Evans}, {Eyer}, {Guerra}, {Hutton}, {Jordi},
  {Klioner}, {Lammers}, {Lindegren}, {Luri}, {Mignard}, {Panem}, {Pourbaix},
  {Randich}, {Sartoretti}, {Soubiran}, {Tanga}, {Walton}, {Bailer-Jones},
  {Bastian}, {Drimmel}, {Jansen}, {Katz}, {Lattanzi}, {van Leeuwen}, {Bakker},
  {Cacciari}, {Casta{\~n}eda}, {De Angeli}, {Fabricius}, {Fouesneau},
  {Fr{\'e}mat}, {Galluccio}, {Guerrier}, {Heiter}, {Masana}, {Messineo},
  {Mowlavi}, {Nicolas}, {Nienartowicz}, {Pailler}, {Panuzzo}, {Riclet}, {Roux},
  {Seabroke}, {Sordo}, {Th{\'e}venin}, {Gracia-Abril}, {Portell}, {Teyssier},
  {Altmann}, {Andrae}, {Audard}, {Bellas-Velidis}, {Benson}, {Berthier},
  {Blomme}, {Burgess}, {Busonero}, {Busso}, {C{\'a}novas}, {Carry}, {Cellino},
  {Cheek}, {Clementini}, {Damerdji}, {Davidson}, {de Teodoro}, {Nu{\~n}ez
  Campos}, {Delchambre}, {Dell'Oro}, {Esquej}, {Fern{\'a}ndez-Hern{\'a}ndez},
  {Fraile}, {Garabato}, {Garc{\'\i}a-Lario}, {Gosset}, {Haigron}, {Halbwachs},
  {Hambly}, {Harrison}, {Hern{\'a}ndez}, {Hestroffer}, {Hodgkin}, {Holl},
  {Jan{\ss}en}, {Jevardat de Fombelle}, {Jordan}, {Krone-Martins}, {Lanzafame},
  {L{\"o}ffler}, {Marchal}, {Marrese}, {Moitinho}, {Muinonen}, {Osborne},
  {Pancino}, {Pauwels}, {Recio-Blanco}, {Reyl{\'e}}, {Riello}, {Rimoldini},
  {Roegiers}, {Rybizki}, {Sarro}, {Siopis}, {Smith}, {Sozzetti}, {Utrilla},
  {van Leeuwen}, {Abbas}, {{\'A}brah{\'a}m}, {Abreu Aramburu}, {Aerts},
  {Aguado}, {Ajaj}, {Aldea-Montero}, {Altavilla}, {{\'A}lvarez}, {Alves},
  {Anders}, {Anderson}, {Anglada Varela}, {Antoja}, {Baines}, {Baker},
  {Balaguer-N{\'u}{\~n}ez}, {Balbinot}, {Balog}, {Barache}, {Barbato},
  {Barros}, {Barstow}, {Bartolom{\'e}}, {Bassilana}, {Bauchet}, {Becciani},
  {Bellazzini}, {Berihuete}, {Bernet}, {Bertone}, {Bianchi}, {Binnenfeld},
  {Blanco-Cuaresma}, {Blazere}, {Boch}, {Bombrun}, {Bossini}, {Bouquillon},
  {Bragaglia}, {Bramante}, {Breedt}, {Bressan}, {Brouillet}, {Brugaletta},
  {Bucciarelli}, {Burlacu}, {Butkevich}, {Buzzi}, {Caffau}, {Cancelliere},
  {Cantat-Gaudin}, {Carballo}, {Carlucci}, {Carnerero}, {Carrasco},
  {Casamiquela}, {Castellani}, {Castro-Ginard}, {Chaoul}, {Charlot}, {Chemin},
  {Chiaramida}, {Chiavassa}, {Chornay}, {Comoretto}, {Contursi}, {Cooper},
  {Cornez}, {Cowell}, {Crifo}, {Cropper}, {Crosta}, {Crowley}, {Dafonte},
  {Dapergolas}, {David}, {David}, {de Laverny}, {De Luise}, \& {De
  March}}]{gaiadr3}
{Gaia Collaboration}, {Vallenari}, A., {Brown}, A.~G.~A., {et~al.} 2023, \aap,
  674, A1

\bibitem[{{Garate-Nu{\~n}ez} {et~al.}(2024){Garate-Nu{\~n}ez}, {Robotham},
  {Bellstedt}, {Davies}, \& {Mart{\'\i}nez-Lombilla}}]{nunez23}
{Garate-Nu{\~n}ez}, L.~P., {Robotham}, A. S.~G., {Bellstedt}, S., {Davies}, L.
  J.~M., \& {Mart{\'\i}nez-Lombilla}, C. 2024, \mnras, 531, 2517

\bibitem[{{Gillis} {et~al.}(2020){Gillis}, {Schrabback}, {Marggraf},
  {Mandelbaum}, {Massey}, {Rhodes}, \& {Taylor}}]{gillis20}
{Gillis}, B.~R., {Schrabback}, T., {Marggraf}, O., {et~al.} 2020, \mnras, 496,
  5017

\bibitem[{{Hasan} \& {Burrows}(1995)}]{hasan95}
{Hasan}, H. \& {Burrows}, C.~J. 1995, \pasp, 107, 289

\bibitem[{{Idiart} {et~al.}(2002){Idiart}, {Michard}, \& {de Freitas
  Pacheco}}]{idiart02}
{Idiart}, T.~P., {Michard}, R., \& {de Freitas Pacheco}, J.~A. 2002, \aap, 383,
  30

\bibitem[{{Infante-Sainz} {et~al.}(2024){Infante-Sainz}, {Akhlaghi}, \&
  {Eskandarlou}}]{raul24}
{Infante-Sainz}, R., {Akhlaghi}, M., \& {Eskandarlou}, S. 2024, Research Notes
  of the American Astronomical Society, 8, 22

\bibitem[{{Infante-Sainz} {et~al.}(2020){Infante-Sainz}, {Trujillo}, \&
  {Rom{\'a}n}}]{infantesainz20}
{Infante-Sainz}, R., {Trujillo}, I., \& {Rom{\'a}n}, J. 2020, MNRAS, 491, 5317

\bibitem[{{Jim{\'e}nez-Teja} {et~al.}(2015){Jim{\'e}nez-Teja}, {Ben{\'\i}tez},
  {Molino}, \& {Fernandes}}]{jimenez15}
{Jim{\'e}nez-Teja}, Y., {Ben{\'\i}tez}, N., {Molino}, A., \& {Fernandes},
  C.~A.~C. 2015, \mnras, 453, 1136

\bibitem[{{King}(1971)}]{king71}
{King}, I.~R. 1971, \pasp, 83, 199

\bibitem[{{Knapen} \& {Trujillo}(2017)}]{knapen17}
{Knapen}, J.~H. \& {Trujillo}, I. 2017, in Astrophysics and Space Science
  Library, Vol. 434, Outskirts of Galaxies, ed. J.~H. {Knapen}, J.~C. {Lee}, \&
  A.~{Gil de Paz}, 255

\bibitem[{{Lanzetta} {et~al.}(2024){Lanzetta}, {Gromoll}, {Shara}, {Berg},
  {Garland}, {Mancini}, {Valls-Gabaud}, {Walter}, \& {Webb}}]{Lanzetta23}
{Lanzetta}, K.~M., {Gromoll}, S., {Shara}, M.~M., {et~al.} 2024, \mnras, 529,
  197

\bibitem[{{Liu} {et~al.}(2022){Liu}, {Abraham}, {Gilhuly}, {van Dokkum},
  {Martin}, {Li}, {Greco}, {Lokhorst}, {Chen}, {Danieli}, {Keim}, {Merritt},
  {Miller}, {Pasha}, {Polzin}, {Shen}, \& {Zhang}}]{liu22}
{Liu}, Q., {Abraham}, R., {Gilhuly}, C., {et~al.} 2022, \apj, 925, 219

\bibitem[{{Liu} {et~al.}(2025){Liu}, {Abraham}, {Martin}, {Bowman}, {Dokkum},
  {Danieli}, {Patel}, {Janssens}, {Shen}, {Chen}, {Karunakaran}, {Keim},
  {Lokhorst}, {Pasha}, \& {Welch}}]{li25}
{Liu}, Q., {Abraham}, R., {Martin}, P.~G., {et~al.} 2025, \apj, 979, 175

\bibitem[{{Liu} {et~al.}(2023){Liu}, {Abraham}, {Martin}, {Bowman}, {van
  Dokkum}, {Janssens}, {Chen}, {Keim}, {Lokhorst}, {Pasha}, {Shen}, \&
  {Zhang}}]{liu23}
{Liu}, Q., {Abraham}, R., {Martin}, P.~G., {et~al.} 2023, \apj, 953, 7

\bibitem[{{L{\'o}pez-Sanjuan} {et~al.}(2024){L{\'o}pez-Sanjuan}, {V{\'a}zquez
  Rami{\'o}}, {Xiao}, {Yuan}, {Carrasco}, {Varela}, {Crist{\'o}bal-Hornillos},
  {Tremblay}, {Ederoclite}, {Mar{\'\i}n-Franch}, {Cenarro}, {Coelho}, {Daflon},
  {del Pino}, {Dom{\'\i}nguez S{\'a}nchez}, {Fern{\'a}ndez-Ontiveros},
  {Hern{\'a}n-Caballero}, {Jim{\'e}nez-Esteban}, {Alcaniz}, {Angulo}, {Dupke},
  {Hern{\'a}ndez-Monteagudo}, {Moles}, \& {Sodr{\'e}}}]{carlinhos23}
{L{\'o}pez-Sanjuan}, C., {V{\'a}zquez Rami{\'o}}, H., {Xiao}, K., {et~al.}
  2024, \aap, 683, A29

\bibitem[{{Marin-Franch} {et~al.}(2015){Marin-Franch}, {Taylor}, {Cenarro},
  {Cristobal-Hornillos}, \& {Moles}}]{Antonio15}
{Marin-Franch}, A., {Taylor}, K., {Cenarro}, J., {Cristobal-Hornillos}, D., \&
  {Moles}, M. 2015, in IAU General Assembly, Vol.~29, 2257381

\bibitem[{{Mart{\'\i}nez-Delgado} {et~al.}(2008){Mart{\'\i}nez-Delgado},
  {Pe{\~n}arrubia}, {Gabany}, {Trujillo}, {Majewski}, \& {Pohlen}}]{martinz08}
{Mart{\'\i}nez-Delgado}, D., {Pe{\~n}arrubia}, J., {Gabany}, R.~J., {et~al.}
  2008, \apj, 689, 184

\bibitem[{{Mart{\'\i}nez-Lombilla} \& {Knapen}(2019)}]{martinexlomnilla19}
{Mart{\'\i}nez-Lombilla}, C. \& {Knapen}, J.~H. 2019, \aap, 629, A12

\bibitem[{{Michard}(2002)}]{michard02}
{Michard}, R. 2002, \aap, 384, 763

\bibitem[{{Middlemass} {et~al.}(1989){Middlemass}, {Clegg}, \&
  {Walsh}}]{middlemass89}
{Middlemass}, D., {Clegg}, R.~E.~S., \& {Walsh}, J.~R. 1989, \mnras, 239, 1

\bibitem[{{Mighell}(2005)}]{mighell05}
{Mighell}, K.~J. 2005, \mnras, 361, 861

\bibitem[{{Mihos} {et~al.}(2013){Mihos}, {Harding}, {Spengler}, {Rudick}, \&
  {Feldmeier}}]{mihos13}
{Mihos}, J.~C., {Harding}, P., {Spengler}, C.~E., {Rudick}, C.~S., \&
  {Feldmeier}, J.~J. 2013, \apj, 762, 82

\bibitem[{{Moffat}(1969)}]{moffat69}
{Moffat}, A.~F.~J. 1969, \aap, 3, 455

\bibitem[{{Nardiello} {et~al.}(2022){Nardiello}, {Bedin}, {Burgasser},
  {Salaris}, {Cassisi}, {Griggio}, \& {Scalco}}]{nardiello22}
{Nardiello}, D., {Bedin}, L.~R., {Burgasser}, A., {et~al.} 2022, \mnras, 517,
  484

\bibitem[{{Peters} {et~al.}(2017){Peters}, {van der Kruit}, {Knapen},
  {Trujillo}, {Fliri}, {Cisternas}, \& {Kelvin}}]{peters17}
{Peters}, S.~P.~C., {van der Kruit}, P.~C., {Knapen}, J.~H., {et~al.} 2017,
  \mnras, 470, 427

\bibitem[{{Rom{\'a}n} {et~al.}(2020){Rom{\'a}n}, {Trujillo}, \&
  {Montes}}]{roman20}
{Rom{\'a}n}, J., {Trujillo}, I., \& {Montes}, M. 2020, \aap, 644, A42

\bibitem[{{Sandin}(2014)}]{sandin14}
{Sandin}, C. 2014, \aap, 567, A97

\bibitem[{{Sandin}(2015)}]{sandin15}
{Sandin}, C. 2015, \aap, 577, A106

\bibitem[{{Sedighi} {et~al.}(2025){Sedighi}, {Sharbaf}, {Trujillo},
  {Eskandarlou}, {Golini}, {Infante-Sainz}, {Raji}, {Zaritsky}, {Ardakani},
  {Chamba}, {Hosseini-ShahiSavandi}, {Donnerstein}, {D'Onofrio}, {Martin},
  {Montes}, \& {Rom{\'a}n}}]{nafise25}
{Sedighi}, N., {Sharbaf}, Z., {Trujillo}, I., {et~al.} 2025, The Open Journal
  of Astrophysics, 8, 73

\bibitem[{{Slater} {et~al.}(2009){Slater}, {Harding}, \& {Mihos}}]{slater09}
{Slater}, C.~T., {Harding}, P., \& {Mihos}, J.~C. 2009, \pasp, 121, 1267

\bibitem[{{Trujillo} {et~al.}(2001{\natexlab{a}}){Trujillo}, {Aguerri}, {Cepa},
  \& {Guti{\'e}rrez}}]{nacho01a}
{Trujillo}, I., {Aguerri}, J.~A.~L., {Cepa}, J., \& {Guti{\'e}rrez}, C.~M.
  2001{\natexlab{a}}, \mnras, 321, 269

\bibitem[{{Trujillo} {et~al.}(2001{\natexlab{b}}){Trujillo}, {Aguerri}, {Cepa},
  \& {Guti{\'e}rrez}}]{nacho01b}
{Trujillo}, I., {Aguerri}, J.~A.~L., {Cepa}, J., \& {Guti{\'e}rrez}, C.~M.
  2001{\natexlab{b}}, \mnras, 328, 977

\bibitem[{{Trujillo} \& {Fliri}(2016)}]{trujillo16}
{Trujillo}, I. \& {Fliri}, J. 2016, \apj, 823, 123

\bibitem[{{van Dokkum} {et~al.}(2019){van Dokkum}, {Gilhuly}, {Bonaca},
  {Merritt}, {Danieli}, {Lokhorst}, {Abraham}, {Conroy}, \&
  {Greco}}]{vandokkum19}
{van Dokkum}, P., {Gilhuly}, C., {Bonaca}, A., {et~al.} 2019, \apjl, 883, L32

\bibitem[{{Watkins} {et~al.}(2024){Watkins}, {Kaviraj}, {Collins}, {Knapen},
  {Kelvin}, {Duc}, {Rom{\'a}n}, \& {Mihos}}]{watkins24}
{Watkins}, A.~E., {Kaviraj}, S., {Collins}, C.~C., {et~al.} 2024, \mnras, 528,
  4289

\bibitem[{{Xin} {et~al.}(2018){Xin}, {Ivezi{\'c}}, {Lupton}, {Peterson},
  {Yoachim}, {Jones}, {Claver}, \& {Angeli}}]{xin18}
{Xin}, B., {Ivezi{\'c}}, {\v{Z}}., {Lupton}, R.~H., {et~al.} 2018, \aj, 156,
  222

\end{thebibliography}
